\documentclass[
superscriptaddress,
preprint,
prd,tightenlines,nofootinbib,
eqsecnum]{revtex4-2}

\usepackage{amsfonts,amsmath,amssymb}
\usepackage[usenames,dvipsnames]{xcolor}
\usepackage{bm}
\usepackage{graphicx} 
\usepackage[colorlinks]{hyperref}
\usepackage{mathrsfs}
\usepackage{multirow}
\usepackage{tensor}
\usepackage{tabularx}
\newcolumntype{Y}{>{\centering\arraybackslash}X}
\AtBeginDocument{\usepackage{booktabs}}

\definecolor{darkgreen}{rgb}{0,0.5,0}

\hypersetup{
    bookmarks=true,         
    unicode=false,          
    pdftoolbar=true,        
    pdfmenubar=true,        
    pdffitwindow=false,     
    pdfstartview={FitH},    
    pdftitle={My title},    
    pdfauthor={Author},     
    pdfsubject={Subject},   
    pdfcreator={Creator},   
    pdfproducer={Producer}, 
    pdfkeywords={keyword1} {key2} {key3}, 
    pdfnewwindow=true,      
    colorlinks=true,       
    linkcolor=red,          
    citecolor=cyan,        
    filecolor=magenta,      
    urlcolor=darkgreen,           
    linktocpage=true
}

\newcommand{\di}{\mathrm{i}}

\newcommand{\ui}{\mathrm{i}}

\newcommand{\be}{\begin{equation}}
\newcommand{\ee}{\end{equation}}

\newcommand{\nn}{\nonumber}
\newcommand{\dd}{\mathrm{d}}

\usepackage{ulem}
\newcommand{\tmass}{M}
\newcommand{\ADM}{\mathcal{M}}

\allowdisplaybreaks

\usepackage{etoolbox}
\makeatletter
\patchcmd{\frontmatter@abstract@produce}
  {\vskip200\p@\@plus1fil
   \penalty-200\relax
   \vskip-200\p@\@plus-1fil}
\makeatother

\begin{document}

\title{Complete gravitational-waveform amplitude modes for quasi-circular compact binaries to the 3.5PN order}

\author{Quentin \textsc{Henry}}\email{quentin.henry@aei.mpg.de}
\affiliation{Max Planck Institute for Gravitational Physics\\ (Albert Einstein Institute), D-14476 Potsdam, Germany}

\date{\today}

\begin{abstract}
We complete the computation of the gravitational waveform amplitude for non-spinning compact binaries to the third and a half post-Newtonian (PN) order in the quasi-circular case. This is done by performing a spin-weighted spherical harmonics decomposition of the amplitude. This computation is achieved using the post-Newtonian-Multipolar-post-Minkowskian formalism. The amplitude modes are written in a PN expanded way as well as a factorized form suitable for EOB template building. Combining this work with previous ones, we provide in the supplemental materials the expressions of the modes including all non-spinning and spinning effects to the 3.5PN order for non-precessing quasi-circular binaries.
\end{abstract}


\maketitle


\section{Introduction}\label{sec:intro}

The gravitational wave (GW) detections by current detectors LIGO-Virgo-KAGRA \cite{LVCO1O2,catalogO3a,catalogO3b}, as well as next generation detectors, such as LISA~\cite{LISA17} and Einstein Telescope~\cite{ET10} require precise modeling of the waveform in order to extract the signal from the noise. Although only numerical relativity can provide accurate waveforms close to the merger, several analytical methods produce theoretical predictions of the inspiral phase. To model this phase, three main types of expansion exist. First of all, the post-Newtonian (PN) one consists in performing a Taylor expansion of the equations using the small parameter $v^2/c^2$, where $v$ is the relative velocity of the two companions of a binary system and $c$ the speed of light. In the standard nomenclature, a $n$PN quantity is of order $\mathcal{O}(1/c^{2n})$. The second one is the post-Minkowskian (PM) expansion, in which we consider the small parameter $G M/rc^2$, where $G$ is Newton's constant, $M$ the total mass of the system and $r$ the separation. The last one consists in performing an expansion using the small parameter is the mass ratio $q=m_1/m_2$. Different frameworks using these types of approximations provide analytical solutions for physical observables such as the radiated energy flux or the GW amplitude and phase. These partial results overlap and can be combined to numerical relativity notably using the effective-one-body (EOB) formalism~\cite{BDEOB99,BuonD00} which produces accurate waveforms over the entire parameter space.

In the radiative sector, the energy flux as well as the GW phase are known to the 3.5PN order beyond the Einstein quadrupole formula \cite{BFIJ02,BDEI04} and are now being pushed to 4PN~\cite{MHLMFB20,Jij3PN,DDRIR1,DDRIR2,BFL22,TLB22}. However, only partial results are currently available for the amplitude of the gravitational waveform to the 3.5PN order. The amplitude can be computed in the form of a spin-weighted spherical harmonics mode decomposition, parametrised as usual by two numbers $(\ell,m)$.

The test-mass limit approximation provides extremely useful results in the amplitude modes at the leading order in the mass ratio. In the parameter space where both the PN and the test-mass limit approximations are valid, these results should agree, which constitutes a very powerful check. The amplitude modes in the test-mass limit are displayed in, e.g.~\cite{TSasa94,TTS96,FI10}.

For non-spinning binaries, the amplitude modes in the PN approximation have been derived consistently to the 3PN order~\cite{BFIS08}. Regarding the 3.5PN accuracy, the $(\ell,m)=(2,2)$ mode was published in~\cite{FMBI12}, the (3,3) and (3,1) modes in~\cite{FBI15} and the (2,1) mode in~\cite{Jij3PN}. Some other modes were derived in~\cite{F09}, especially some for high $\ell$ at the relative 1PN order, matching the 3.5PN accuracy in the waveform. These modes have been used for EOB template building. It has been found that factorizing their expressions in several blocks increases the convergence of the EOB model when compared to numerical relativity~\cite{Damour:2007yf}. Thus, the PN expanded modes at the third PN order were factorized in, e.g. \cite{DIN09,Pan:2011gk} to give more accurate EOB templates. Although the factorized form of the (2,2), (3,3) and (3,1) modes are known for the 3.5PN accuracy in the waveform, the one of the (2,1) mode was never published. Furthermore, it was found out recently that an error in a published version of the (2,1) mode at the 2.5PN order propagated in the EOB models (see~\cite{HMK22} for justification). The PN expanded form of the (2,1) mode eventually got corrected in the erratum of~\cite{BFIS08} but the mistake in the EOB models remained at the time of publication of this work.

In order to prevent new propagation of mistakes and to avoid multiplying references for different terms in the PN expanded modes as well as their factorized version, this paper aims at completing the full gravitational waveform to the 3.5PN order, including spin effects for a quasi-circular and non-precessing motion. The spin effects in the waveform modes to this order were derived in the recent work~\cite{HMK22} and we now complete the missing non-spinning coefficients. We also write the modes in a convenient way for EOB template building. Due to the length of the expressions, the spin contributions to the amplitude modes are not be displayed here but can be found in~\cite{HMK22}. However, the Supplemental Materials attached to this paper contain both non-spinning and spinning coefficients to the 3.5PN order~\cite{SuppMaterial}.

In the present paper, we use the post-Newtonian-Multipolar-post-Minkowskian (PN-MPM) formalism; see~\cite{BlanchetLR} for a detailed review of this approach. This method combines the PN and the PM expansions, so that we consider the small parameters $v^2/c^2 \sim G\tmass/c^2r \ll 1$. We will qualitatively describe this method in Sec.~\ref{subsec:computationalmethod}.

The paper is organized as follows. In Sec.~\ref{sec:summaryformalism}, we define the spherical modes and recall their relation with the radiative multipole moments. We also explain qualitatively how to derive the radiative moments within the PN-MPM scheme. We refer to previous works for technical details of this derivation. In Sec.~\ref{sec:results}, we give the results for the non-spinning terms in amplitude modes, written in a conventional Taylor expansion form in Sec.~\ref{subsec:hlm}, and factorized conveniently for EOB usage in Sec.~\ref{subsec:eob}. Appendix \ref{app:memory} contains generalized formulas to compute some particular hereditary effects. Appendix \ref{app:eob} contains the different quantities appearing in the factorized modes.
We also provide our results for the full waveform modes including non-spinning and spinning effects to the 3.5PN order as \emph{Mathematica} files in the Supplemental Materials~\cite{SuppMaterial}.


\section{Brief overview of the method}\label{sec:summaryformalism}
\subsection{Spherical harmonics decomposition of the gravitational field}

The transverse-traceless (TT) projection $h_{ij}^\text{TT}$ of the gravitational field can be uniquely decomposed in terms of a set of symmetric trace-free (STF) multipole moments $U_L$ and $V_L$, called radiative multipole moments, as \cite{Th80}\footnote{We use the signature $(-,+,+,+)$. Latin indices stand for spatial coordinates, i.e.\ $i=1,2,3$ and the multi-index notation $L=i_1\dots i_\ell$. The weighted symmetrization operator is noted by parenthesis around indices.}
\begin{align}\label{eq:hij}
h_{ij}^\text{TT} &= \frac{4G}{c^2R} \,\mathcal{P}_{ijkl} (\bm{N}) \sum^{+\infty}_{\ell=2}\frac{1}{c^\ell \ell !} \left\{ N_{L-2} \,U_{klL-2}(T_R) - \frac{2\ell}{c(\ell+1)} \,N_{aL-2} \,\varepsilon_{ab(k} \,V_{l)bL-2}(T_R)\right\} +\mathcal{O}\left( \frac{1}{R^2} \right) \,,
\end{align}
where $R$ is the distance between the source and the observer, $\bm{N}$ is the direction of propagation of the GW and $T_R= T- R/c$ is the retarded time in some radiative gauge in which $T_R$ is asymptotically null. The quantity $\mathcal{P}_{ijkl} = \mathcal{P}_{i(k}\mathcal{P}_{l)j}-\frac{1}{2}\mathcal{P}_{ij}\mathcal{P}_{kl}$ is the TT projection operator, where $\mathcal{P}_{ij}=\delta_{ij}-N_iN_j$. The polarization waveforms are defined by
\begin{subequations}\label{eq:hpluscross}
\begin{align}
h_+ &= \frac{1}{2}(P_i P_j - Q_i Q_j)h_{ij}^\text{TT},\\
h_\times &= \frac{1}{2}(P_i Q_j + Q_i P_j)h_{ij}^\text{TT},
\end{align}
\end{subequations}
where the vectors $(\bm{P},\bm{Q},\bm{N})$ form an orthonormal triad properly defined in, e.g. Sec. II. A. of~\cite{HMK22}. As usual, we decompose $h_+ -\di h_\times$ in a spin-weighted spherical harmonics basis of weight -2 \cite{K07}
\begin{equation}\label{eq:h}
h\equiv h_+ -\di h_\times = \sum_{l=0}^\infty \sum_{m=-\ell}^{\ell} h_{\ell m} Y^{\ell m}_{-2}(\Theta,\Phi),
\end{equation}
where the two angles $(\Theta,\Phi)$ characterize the direction of propagation $\bm{N}$. The amplitude modes $h_{\ell m}$ are then linked to the radiative multipole moments through
\begin{equation}\label{eq:hlm}
h_{\ell m} = -\frac{G}{\sqrt{2} R c^{\ell+2}}\left( U^{\ell m} -\frac{\di}{c} V^{\ell m} \right).
\end{equation}
In our convention, $U^{\ell m}$ and $V^{\ell m}$ are given by\footnote{Note that the choice of definitions in Ref.~\cite{K07,Pan11} on $h_{\ell m}$ differs from ours by a global minus sign for each $\ell$ and $m$ due to a different definition of the vector basis $(\bm{P},\bm{Q},\bm{N})$.}
\begin{subequations}\label{eq:UlmVlm}
\begin{align}
U^{\ell m} &= \frac{4}{\ell!}\,\sqrt{\frac{(\ell+1)(\ell+2)}{2\ell(\ell-1)}}\,\alpha_L^{\ell m}\,U_L\,,\\
V^{\ell m} &= -\frac{8}{\ell!}\,\sqrt{\frac{\ell(\ell+2)}{2(\ell+1)(\ell-1)}}\,\alpha_L^{\ell m}\,V_L\,,
\end{align}
\end{subequations}
with $\alpha_L^{\ell m} \equiv \int \dd \Omega\,\hat{N}_L\,\overline{Y}^{\,\ell m}$ being defined from the complex conjuguate of the ordinary spherical harmonics $Y^{\ell m}$. Its explicit expression is displayed in Eq.~(4.7) of Ref.~\cite{Jij3PN}.

As we can see from Eqs. \eqref{eq:hlm}-\eqref{eq:UlmVlm}, one needs to derive the radiative multipole moments $U_L$ and $V_L$ to consistent order to obtain the amplitude modes $h_{\ell m}$. The derivation of these moments are described in the following section.

\subsection{Computational method}\label{subsec:computationalmethod}

In this section, we describe qualitatively how to derive the radiative multipole moments required for the knowledge of the full waveform at the 3.5PN order. We will not give the technical details of the intermediate computations as they were broadly described in other papers cited below. A more detailed overview of each step can be found in, e.g. Sec. III. of~\cite{HMK22}.\\

The PN-MPM formalism provides an efficient way to compute the radiative multipole moments defined in~\eqref{eq:hij}, see Ref.~\cite{BlanchetLR} for a review. The first step of this computation consists in deriving the equations of motion (EOM) of the individual compact bodies. They are displayed to the 3.5PN order in \cite{FMBI12} in harmonic coordinates and have been derived more recently to the 4PN order in \cite{MBBF17} in the same gauge. In the case of quasi-circular orbits, one can extract the orbital frequency of the system $\omega$ from the EOM which allows defining the orbital phase through
\begin{equation}\label{eq:phi}
\phi = \int \dd t\, \omega(t).
\end{equation}
To derive the phase to a given PN order, one needs to know both the conserved energy and the radiated energy flux at the same order. As mentionned in the introduction, this phase is known to the 3.5PN order. However, the PN precision required on the radiative multipole moments to derive the radiated flux, hence the phase, is lower than the one required to derive the amplitude modes. In other words, one needs more PN information to derive the modes than the energy flux at a given PN order.

In this project, we computed additional PN terms in the $U_L$ for $\ell\in\{4,\dots,9\}$ and $V_L$ for $\ell\in\{3,\dots,8\}$. The mass quadrupole, mass octupole and current quadrupole were already obtained in previous works to consistent order \cite{FMBI12,FBI15,Jij3PN}.

For this project, we first computed the so-called source multipole moments. They are defined by constructing the most general solution of Einstein's equations in vacuum in the form of a multipolar post-Minkowskian expansion outside the system. They are then matched to the PN metric, valid in a spatial zone near the system. By doing so, we find that the source multipole moments are integrals over quantities that contain the stress-energy tensor and the PN metric. The explicit expressions of the source moments are given in, e.g. Eqs. (2.31) and (2.33) of~\cite{Jij3PN}. The metric as well as the stress-energy tensor were already known from previous works, see e.g.~\cite{MHLMFB20}, in which these quantities were derived for the 4PN accuracy of the source mass quadrupole.

In order to perform this matching between different representations of the gravitational field, we must resort to a regularisation scheme. We used in this paper the \textit{Hadamard partie finie} regularisation. However, it was found out that this regularisation starts to fail at high PN orders. In fact, only dimensional regularisation has been proved to be satisfactory. It was shown in previous works that for non-spinning binaries, these two regularisations start to differ to the 3PN order \cite{BDE04,BDEI04,BDEI05dr,GRoss10,DDRIR1,DDRIR2}. In this project, the missing multipoles to complete the full non-spinning waveform at the 3.5PN order were at most required to the 2.5PN order so that we could use the \textit{Hadamard partie finie} regularisation safely. The integration methods for this regularisation scheme are provided in, e.g. \cite{BFP98}.

This integration leads to the expression of the source multipole moments in an arbitrary frame for general orbits. The next step consists in reducing the expressions to the center of mass (CM) frame and impose a quasi-circular motion. To do so, we use the CM position displayed in, e.g. \cite{BBFM17} and neglect eccentricities. 

As a last step, we deduce the radiative moments from the source moments. Due to the non-linearity of the gravitational field, they differ by some non-linear interactions of different moments. We find two main types of contributions: the so-called instantaneous interactions, which correspond to at least quadratic products of source multipoles evaluated at retarded time; and hereditary interactions which contain for example tail and memory integrals over the past history of the source preceding the retarded time. The general expressions of the interactions were derived entirely to the 3.5PN order in \cite{FBI15}. By using these general formulas and the integration methods presented in Sec. VII. of \cite{BFIS08}, we obtained all the radiative multipole moments that were needed to compute the full waveform to the 3.5PN order. In Appendix~\ref{app:memory}, we give the generalization to higher PN order of some integrals required for computing the memory interactions.

Finally, we inserted the values of these moments in Eqs.~\eqref{eq:hlm}-\eqref{eq:UlmVlm} to obtain the GW modes displayed in the following section.

\section{Results}\label{sec:results}

In this section, we display the results for the waveform modes written in two different ways. In Sec.~\ref{subsec:hlm}, we give their expressions in the conventional PN expanded way in the form of, e.g. \cite{BFIS08}. In Sec.~\ref{subsec:eob}, we write them in a factorized form convenient for the EOB method.

\subsection{Spin-weighted spherical modes}\label{subsec:hlm}

The amplitude modes defined in Eq.~\eqref{eq:hlm} can be written in terms of the phase variable $\phi$ defined in~\eqref{eq:phi}. In order to simplify the explicit expressions of the modes, we introduce the new phase variable
\begin{equation}\label{psi}
\psi \equiv \phi - \frac{2G \ADM \omega}{c^3} \ln\left(\frac{\omega}{\omega_0}\right),
\end{equation}
where $\ADM$ is the ADM mass and where the scale constant $\omega_0$ is introduced in the definition of the tail integrals~\cite{BFIS08}. This redefinition absorbs most of the logarithms of the orbital frequency in the oscillatory part of the modes. The GW modes appearing in~\eqref{eq:h} read
\begin{align}\label{eq:hlmexpl}
	h_{\ell m} = \frac{2 G \tmass \,\nu \,x}{R \,c^2} \,
	\sqrt{\frac{16\pi}{5}}\,\hat{H}_{\ell m}\,e^{-\di m \psi}\,,
\end{align}
where $\tmass$ is the total mass of the system, $\nu=m_1 m_2/M^2$ is the symmetric mass ratio and where we introduced the convenient PN parameter
\begin{equation}
x=\left(\frac{G \tmass\omega}{c^3}\right)^{2/3}.
\end{equation}
The values of the $\hat{H}_{\ell m}$'s for the full waveform produced by binary a binary system of non-spinning compact objects to the 3.5PN order are given by
\begin{subequations} \label{HlmExp}
\begin{align}
 \hat{H}_{22}&=1+x \left(-\frac{107}{42}+\frac{55 \nu }{42}\right)+2 \pi
x^{3/2}+x^2 \left(-\frac{2173}{1512}-\frac{1069 \nu }{216}+\frac{2047 \nu
^2}{1512}\right) \nonumber \\ &+x^{5/2} \left(-\frac{107 \pi }{21}-24 \di \nu
+\frac{34 \pi \nu }{21}\right)+x^3 \bigg(\frac{27027409}{646800}-\frac{856
\gamma_E}{105}+\frac{428 \di \pi }{105}+\frac{2 \pi ^2}{3}\nonumber \\ &
+\left(-\frac{278185}{33264}+\frac{41 \pi^2}{96}\right) \nu -\frac{20261 \nu
^2}{2772}+\frac{114635 \nu^3}{99792}-\frac{428}{105} \ln (16 x)\bigg)\nn \\
& + x^{7/2}\left( -\frac{2173}{756}\pi + \left(-\frac{2495}{378}\pi +\frac{14333}{162}\di \right)\nu + \left(\frac{40}{27}\pi -\frac{4066}{945}\di \right)\nu^2 \right)\label{h22}\,,\\ 
\hat{H}_{21} &=\frac{1}{3} \di \,\delta \bigg[x^{1/2}+x^{3/2} \left(-\frac{17}{28}+\frac{5
\nu }{7}\right)+x^2 \left(\pi +\di \left(-\frac{1}{2}-2 \ln 2\right)\right)
\nonumber \\ & +x^{5/2} \left(-\frac{43}{126}-\frac{509 \nu }{126}+\frac{79
\nu^2}{168}\right)+x^3 \bigg(-\frac{17 \pi }{28}+\frac{3 \pi \nu }{14}
\nonumber \\ &+\di \left(\frac{17}{56}+\nu \left(-\frac{353}{28}-\frac{3 \ln
2}{7}\right)+\frac{17 \ln 2}{14}\right)\bigg) \nn \\
& + x^{7/2}\biggl(\frac{15223771}{1455300}+\frac{\pi^2}{6}-\frac{214}{105}\gamma_\text{E}-\frac{107}{105}\ln(4x)-\ln 2-2(\ln 2)^2\nonumber \\ &+\nu\biggl(-\frac{102119}{2376}+\frac{205}{128}\pi^2\biggr)-\frac{4211}{8316}\nu^2+\frac{2263}{8316}\nu^3+\di \pi\biggl(\frac{109}{210}-2\ln 2\biggr)\biggr)\biggr]\,,\\ 
\hat{H}_{20}& =-\frac{5}{14\sqrt{6}}\left[ 1+ x\left( -\frac{4075}{4032} + \frac{67\nu}{48}\right)  \right] + \mathcal{O}\left(\frac{1}{c^4}\right)\,,\\ 
\hat{H}_{33} &=-\frac{3}{4}\di\sqrt{\frac{15}{14}} \,\delta \bigg[x^{1/2}+x^{3/2} (-4+2 \nu
)+x^2 \left(3 \pi +\di \left(-\frac{21}{5}+6 \ln \left(3/2\right)\right)\right)
\nonumber \\ &+x^{5/2} \left(\frac{123}{110}-\frac{1838 \nu }{165}+\frac{887
\nu ^2}{330}\right)+x^3 \bigg(-12 \pi +\frac{9 \pi \nu }{2} \nonumber \\ &+\di
\left(\frac{84}{5}-24 \ln \left(3/2\right)+\nu \left(-\frac{48103}{1215}+9 \ln
\left(3/2\right)\right)\right)\bigg)\bigg]\nn \\ & + x^{7/2}\biggl(\frac{19388147}{280280} +
  \frac{492}{35} \ln \left(3/2\right) -18\ln^2 (3/2) -
  \frac{78}{7}\gamma_\text{E} + \frac{3}{2} \pi^2 + 6 \ui \pi
  \left(-\frac{41}{35} + 3 \ln (3/2) \right) \nonumber \\ 
&  + \frac{\nu}{8} \left(- \frac{7055}{429} + \frac{41}{8} \pi^2 \right) -
  \frac{318841}{17160} \nu^2 + \frac{8237}{2860} \nu^3 - \frac{39}{7}
  \ln (16x) \biggr)\bigg]\\ 
\hat{H}_{32} &=\frac{1}{3}
\sqrt{\frac{5}{7}} \bigg[x (1-3 \nu )+x^2 \left(-\frac{193}{90}+\frac{145 \nu
}{18}-\frac{73 \nu ^2}{18}\right)+x^{5/2} \left(2 \pi -6 \pi \nu +\di
\left(-3+\frac{66 \nu }{5}\right)\right) \nonumber \\ &+x^3
\left(-\frac{1451}{3960}-\frac{17387 \nu }{3960}+\frac{5557
\nu^2}{220}-\frac{5341 \nu^3}{1320}\right)+ x^{7/2}\left(\frac{193}{30}\di-\frac{193}{45}\pi  \nn \right. \\
& \left. + \nu \left(-\frac{258929}{5400}\di+\frac{136}{9}\pi\right) + \nu^2\left( \frac{33751}{450}\di-\frac{46}{9}\pi \right)\right)\bigg]\,,\\ 
\hat{H}_{31} &=\frac{\di \,\delta}{12\sqrt{14}} \bigg[x^{1/2}+x^{3/2} \left(-\frac{8}{3}-\frac{2 \nu
}{3}\right)+x^2 \left(\pi +\di \left(-\frac{7}{5}-2 \ln 2\right)\right)
\nonumber \\ &+x^{5/2} \left(\frac{607}{198}-\frac{136 \nu }{99}-\frac{247
\nu^2}{198}\right)+x^3 \bigg(-\frac{8 \pi }{3}-\frac{7 \pi \nu }{6} \nonumber
\\ &+\di \left(\frac{56}{15}+\frac{16 \ln 2}{3}+\nu \left(-\frac{1}{15}+\frac{7
\ln 2}{3}\right)\right)\bigg)\bigg] \nn \\ &+ x^{7/2}\biggl( \frac{10753397}{1513512} - 2 \ln 2
  \left( \frac{212}{105} + \ln 2\right) - \frac{26}{21} \gamma_\text{E} +
  \frac{\pi^2}{6} -2 \ui \pi \left( \frac{41}{105} + \ln 2 \right)
  \nonumber \\ & \qquad + \frac{\nu}{8} \bigg(- \frac{1738843}{19305}
  + \frac{41}{8} \pi^2 \bigg) + \frac{327059}{30888} \nu^2 -
  \frac{17525}{15444} \nu^3 - \frac{13}{21} \ln x \biggr)\bigg]\,,\\
\hat{H}_{30} &=-\frac{2}{5} \di \sqrt{\frac{6}{7}} \nu \left[ x^{5/2} +x^{7/2}\left( -\frac{5017}{1296} -\frac{25}{108}\nu  \right)\right] \,,\\
\hat{H}_{44} &=-\frac{8}{9} \sqrt{\frac{5}{7}} \bigg[x (1-3 \nu )+x^2
\left(-\frac{593}{110}+\frac{1273 \nu }{66}-\frac{175 \nu^2}{22}\right)
\nonumber \\ &+x^{5/2} \left(4 \pi -12 \pi \nu +\di \left(-\frac{42}{5}+\nu
\left(\frac{1193}{40}-24 \ln 2\right)+8 \ln 2\right)\right) \nonumber \\ &+x^3
\left(\frac{1068671}{200200}-\frac{1088119 \nu }{28600}+\frac{146879 \nu
^2}{2340}-\frac{226097 \nu^3}{17160}\right) \nn \\
& + x^{7/2}\left( \frac{12453}{275}\di -\frac{1186}{55}\pi-\frac{2372}{55}\di \ln 2 + \nu\left( -\frac{31525499}{140800}\di+\frac{2480}{33}\pi +\frac{4960}{33}\di \ln 2 \right) \nn \right. \\
& \left. + \nu^2\left( \frac{4096237}{21120}\di-\frac{284}{11}\pi -\frac{568}{11}\di \ln 2 \right)\right)\bigg]\,,\\ 
\hat{H}_{43} &=-\frac{9 \di
\,\delta}{4 \sqrt{70}} \bigg[x^{3/2} (1-2 \nu )+x^{5/2}
\left(-\frac{39}{11}+\frac{1267 \nu }{132}-\frac{131 \nu^2}{33}\right)
\nonumber \\ &+x^3 \left(3 \pi -6 \pi \nu +\di \left(-\frac{32}{5}+\nu
\left(\frac{16301}{810}-12 \ln \left(3/2\right)\right)+6 \ln
\left(3/2\right)\right)\right) \nn \\
& + x^{7/2}\left( \frac{7206}{5005}-\frac{82869}{5720}\nu+\frac{104839}{3432}\nu^2-\frac{2987}{572}\nu^3 \right)\bigg]\,,\\ \hat{H}_{42} &=\frac{1}{63}
\sqrt{5} \bigg[x (1-3 \nu )+x^2 \left(-\frac{437}{110}+\frac{805 \nu
}{66}-\frac{19 \nu^2}{22}\right)+x^{5/2} \bigg(2 \pi -6 \pi \nu \nonumber \\ &
+\di \left(-\frac{21}{5}+\frac{84 \nu }{5}\right)\bigg) +x^3
\left(\frac{1038039}{200200}-\frac{606751 \nu }{28600}+\frac{400453 \nu
^2}{25740}+\frac{25783 \nu^3}{17160}\right)\nn \\
& + x^{7/2}\left( \frac{9177}{550}\di - \frac{437}{55}\pi +\nu\left(-\frac{83029}{880}\di + \frac{772}{33}\pi \right)+\nu^2\left(\frac{93081}{1100}\di + \frac{14}{11}\pi \right) \right)\bigg]\,,\\ 
\hat{H}_{41} &=\frac{\di \,\delta}{84\sqrt{10}} \bigg[x^{3/2} (1-2 \nu )+x^{5/2} \left(-\frac{101}{33}+\frac{337
\nu }{44}-\frac{83 \nu^2}{33}\right) \nonumber \\ & +x^3 \left(\pi -2 \pi \nu
+\di \left(-\frac{32}{15}-2 \ln 2+\nu \left(\frac{1661}{30}+4 \ln
2\right)\right)\right)\nn \\
&+ x^{7/2}\left(\frac{42982}{15015}-\frac{513989}{51480}\nu+\frac{196957}{10296}\nu^2-\frac{1195}{572}\nu^3 \right)\bigg]\,,\\
\hat{H}_{40} &=-\frac{1}{504 \sqrt{2}}\left[ 1+x\left( -\frac{180101}{29568}+\frac{27227}{1056}\nu \right)  \right] + \mathcal{O}\left(\frac{1}{c^4}\right)\,,\\ \hat{H}_{55} &=\frac{625 \di
\,\delta}{96 \sqrt{66}} \bigg[x^{3/2} (1-2 \nu )+x^{5/2}
\left(-\frac{263}{39}+\frac{688 \nu }{39}-\frac{256 \nu^2}{39}\right)
\nonumber \\ &+x^3 \left(5 \pi -10 \pi \nu +\di \left(-\frac{181}{14}+\nu
\left(\frac{105834}{3125}-20 \ln \left(5/2\right)\right)+10 \ln
\left(5/2\right)\right)\right)\nn \\
&+ x^{7/2}\left(\frac{9185}{819}-\frac{188765}{3276}\nu+\frac{54428}{819}\nu^2-\frac{10567}{819}\nu^3 \right)\bigg]\,,\\ 
\hat{H}_{54} &=-\frac{32}{9\sqrt{165}} \bigg[x^2 \left(1-5 \nu +5 \nu^2\right)+x^3
\left(-\frac{4451}{910}+\frac{3619 \nu }{130}-\frac{521 \nu ^2}{13}+\frac{339
\nu^3}{26}\right) \nn \\
&+ x^{7/2}\left( -\frac{52}{5}\di +4\pi +8\di \ln 2 +\nu\left(\frac{3351011}{53760}\di -20\pi -40\di \ln 2\right) \nn \right. \\
&\left. +\nu^2\left(-\frac{10923}{128}\di +20\pi +40\di \ln 2\right) \right)\bigg]\,,\\
\hat{H}_{53} &=-\frac{9}{32}\di\sqrt{\frac{3}{110}} \,\delta \bigg[x^{3/2}
(1-2 \nu )+x^{5/2} \left(-\frac{69}{13}+\frac{464 \nu }{39}-\frac{88 \nu
^2}{39}\right)\nonumber \\ & +x^3 \left(3 \pi -6 \pi \nu +\di
\left(-\frac{543}{70}+\nu \left(\frac{83702}{3645}-12 \ln
\left(3/2\right)\right)+6 \ln \left(3/2\right)\right)\right) \nn \\
& + x^{7/2}\left(\frac{12463}{1365}-\frac{56969}{1820}\nu+\frac{2172}{91}\nu^2-\frac{365}{273}\nu^3 \right)\bigg]\,,\\ 
\hat{H}_{52} &=\frac{2}{27\sqrt{55}} \bigg[x^2 \left(1-5 \nu +5 \nu^2\right)+x^3
\left(-\frac{3911}{910}+\frac{3079 \nu }{130}-\frac{413 \nu ^2}{13}+\frac{231
\nu^3}{26}\right)\nn \\
& + x^{7/2}\left( -\frac{26}{5}\di+2\pi+\nu\left( \frac{16237}{336}\di-10\pi \right)+\nu^2\left(-\frac{1861}{20}\di +10\pi  \right) \right)\bigg]\,,\\
\hat{H}_{51} &=\frac{\di \,\delta}{288 \sqrt{385}} \bigg[x^{3/2} (1-2 \nu
)+x^{5/2} \left(-\frac{179}{39}+\frac{352 \nu }{39}-\frac{4 \nu
^2}{39}\right)\nonumber \\ & +x^3 \left(\pi -2 \pi \nu +\di
\left(-\frac{181}{70}-2 \ln 2+\nu \left(\frac{626}{5}+4 \ln
2\right)\right)\right) \nn \\
& +x^{7/2}\left(\frac{5023}{585}-\frac{49447}{2340}\nu+\frac{68}{9}\nu^2+\frac{287}{117}\nu^3 \right)\bigg]\,,\\
\hat{H}_{50} &= \frac{4117\di \nu}{7560\sqrt{462}}x^{7/2}\left[1-\frac{21588}{4117}\nu \right]\,,\\ 
\hat{H}_{66}&=\frac{54}{5 \sqrt{143}} \bigg[x^2 \left(1-5 \nu +5 \nu^2\right)+x^3
\left(-\frac{113}{14}+\frac{91 \nu }{2}-64 \nu^2+\frac{39 \nu
^3}{2}\right)\nn \\
& +x^{7/2}\left(-\frac{249}{14}\di+6\pi+12\di \ln 3 +\nu\left(\frac{21787499}{217728}\di-30\pi-60\di \ln 3 \right) \nn \right. \\
&\left. +\nu^2\left(-\frac{323903}{2592}\di+30\pi+60\di \ln 3 \right) \right)\bigg]\,,\\ 
\hat{H}_{65} &=\frac{3125\di \,\delta}{504 \sqrt{429}} \bigg[x^{5/2}\left(1-4 \nu +3 \nu^2 \right) + x^{7/2}\left(-\frac{149}{24} +\frac{349}{12}\nu-\frac{409}{12}\nu^2+\frac{29}{3}\nu^3 \right)\bigg]\,,\\ 
\hat{H}_{64}
&=-\frac{128}{495} \sqrt{\frac{2}{39}} \bigg[x^2 \left(1-5 \nu +5
\nu^2\right)+x^3 \left(-\frac{93}{14}+\frac{71 \nu }{2}-44 \nu^2+\frac{19 \nu
^3}{2}\right)\nn \\
& +x^{7/2}\left(-\frac{83}{7}\di+4\pi+8\di \ln 2 +\nu\left(\frac{3432215}{49152}\di-20\pi-40\di \ln 2 \right) \nn \right. \\
&\left. +\nu^2\left(-\frac{382365}{4096}\di+20\pi+40\di \ln 2 \right) \right)\bigg]\,,\\ 
\hat{H}_{63}&=-\frac{81\di\,\delta}{616 \sqrt{65}} \bigg[x^{5/2}\left(1-4 \nu +3 \nu^2\right) +x^{7/2}\left(-\frac{133}{24}+\frac{301}{12}\nu-\frac{329}{12}\nu^2+7\nu^3 \right) \bigg]\,,\\ 
\hat{H}_{62} &=\frac{2}{297\sqrt{65}} \bigg[x^2 \left(1-5 \nu +5 \nu^2\right)+x^3
\left(-\frac{81}{14}+\frac{59 \nu }{2}-32 \nu^2+\frac{7 \nu
^3}{2}\right)\nn \\ 
& + x^{7/2}\left(-\frac{83}{14}\di +2\pi+\nu\left(\frac{799789}{13440}\di -10\pi \right)+\nu^2\left(-\frac{19193}{160}\di +10\pi \right) \right)\bigg]\,,\\ 
\hat{H}_{61}&=\frac{\di \,\delta}{8316 \sqrt{26}} \bigg[x^{5/2}\left(1-4 \nu +3 \nu^2\right) + x^{7/2}\left(-\frac{125}{24}+\frac{277}{12}\nu -\frac{289}{12}\nu^2+\frac{17}{3}\nu^3\right)\bigg]\,,\\ 
\hat{H}_{60} &=\frac{4195x}{1419264\sqrt{273}}\left[1-\frac{3612}{839}\nu \right]+\mathcal{O}\left(\frac{1}{c^4}\right)\,,\\ 
\hat{H}_{77} &=-\frac{16807\di \,\delta}{1440} \sqrt{\frac{7}{858}} \bigg[x^{5/2}\left(1-4 \nu +3 \nu^2\right) + x^{7/2}\left(-\frac{319}{34}+\frac{2225}{51}\nu-\frac{2558}{51}\nu^2+\frac{230}{17}\nu^3 \right)\bigg]\,,\\ 
\hat{H}_{76} &=\frac{81}{35}
\sqrt{\frac{3}{143}} x^3 \bigg[1-7 \nu +14 \nu^2-7 \nu^3\bigg]\,,\\ 
\hat{H}_{75} &=\frac{15625\di \,\delta}{26208 \sqrt{66}} \bigg[x^{5/2}\left(1-4 \nu +3 \nu^2\right) +x^{7/2}\left( -\frac{271}{34}+\frac{1793}{51}\nu-\frac{1838}{51}\nu^2+\frac{134}{17}\nu^3 \right)\bigg]\,,\\ 
\hat{H}_{74} &=-\frac{128x^3}{1365}\sqrt{\frac{2}{33}} \bigg[1-7 \nu +14 \nu^2-7 \nu^3\bigg]\,,\\ 
\hat{H}_{73} &=-\frac{243\di \,\delta}{160160}\sqrt{\frac{3}{2}} \bigg[x^{5/2}\left(1-4 \nu +3 \nu^2\right)+x^{7/2}\left(-\frac{239}{34} +\frac{1505}{51}\nu-\frac{1358}{51}\nu^2+\frac{70}{17}\nu^3\right)\bigg]\,,\\ 
\hat{H}_{72} &=\frac{x^3}{3003 \sqrt{3}} \bigg[1-7
\nu +14 \nu^2-7 \nu^3\bigg]\,,\\ 
\hat{H}_{71} &=\frac{\di\,\delta}{864864 \sqrt{2}} \bigg[x^{5/2}\left(1-4 \nu +3 \nu^2\right)+x^{7/2}\left(-\frac{223}{34}+\frac{1361}{51}\nu-\frac{1118}{51}\nu^2 +\frac{38}{17}\nu^3\right)\bigg]\,,\\ 
\hat{H}_{70} &= \mathcal{O}\left( \frac{1}{c^8} \right)\,,\\
\hat{H}_{88} &=-\frac{16384}{63}
\sqrt{\frac{2}{85085}} x^3 \bigg[1-7 \nu +14 \nu^2-7 \nu^3\bigg]\,,\\ 
\hat{H}_{87} &=-\frac{117649\,\di \, \delta}{5184}\sqrt{\frac{7}{24310}}x^{7/2}\bigg[ 1-6\nu+10\nu^2-4\nu^3 \bigg]\,,\\ 
\hat{H}_{86} &=\frac{243}{35}\sqrt{\frac{3}{17017}} x^3 \bigg[1-7 \nu +14 \nu^2-7 \nu^3\bigg]\,,\\ 
\hat{H}_{85} &=\frac{78125\,\di \, \delta}{36288\sqrt{4862}}x^{7/2}\bigg[ 1-6\nu+10\nu^2-4\nu^3 \bigg] \,,\\ 
\hat{H}_{84} &=-\frac{128}{4095}\sqrt{\frac{2}{187}} x^3 \bigg[1-7 \nu +14 \nu^2-7 \nu ^3\bigg]\,,\\ 
\hat{H}_{83} &= -\frac{81\,\di \, \delta}{5824}\sqrt{\frac{3}{1870}}x^{7/2}\bigg[ 1-6\nu+10\nu^2-4\nu^3 \bigg]\,,\\ 
\hat{H}_{82} &=\frac{x^3}{9009\sqrt{85}} \bigg[1-7 \nu +14 \nu^2-7 \nu^3\bigg]\,,\\ 
\hat{H}_{81} &=\frac{\di \, \delta}{741312\sqrt{238}}x^{7/2}\bigg[ 1-6\nu+10\nu^2-4\nu^3 \bigg] \,,\\ 
\hat{H}_{80} &= \mathcal{O}\left( \frac{1}{c^4} \right)\,,\\
\hat{H}_{99} &= \frac{1594323\,\di \, \delta}{7168\sqrt{20995}}x^{7/2}\bigg[ 1-6\nu+10\nu^2-4\nu^3 \bigg]\,, \\
\hat{H}_{97} &= -\frac{5764801\,\di \, \delta}{1410048\sqrt{1235}}x^{7/2}\bigg[ 1-6\nu+10\nu^2-4\nu^3 \bigg]\,, \\
\hat{H}_{95} &= \frac{390625\,\di \, \delta}{4935168\sqrt{247}}x^{7/2}\bigg[ 1-6\nu+10\nu^2-4\nu^3 \bigg]\,, \\
\hat{H}_{93} &= -\frac{81\,\di \, \delta}{113152}\sqrt{\frac{3}{665}}x^{7/2}\bigg[ 1-6\nu+10\nu^2-4\nu^3 \bigg]\,, \\
\hat{H}_{91} &= \frac{\di \, \delta}{9165312\sqrt{2090}}x^{7/2}\bigg[ 1-6\nu+10\nu^2-4\nu^3 \bigg]\,.
\end{align}
\end{subequations}
The modes written without a remainder are all known up to order $\mathcal{O}(1/c^7)$. Note that some modes for $(\ell,0)$ are written with a lower order remainder. As discussed in \cite{HMK22}, the contributions to these modes only arise from the memory integrals in the mass-type radiative moments. Although the PN-MPM does not allow to compute them to high orders, they have been derived to the 3PN order beyond their leading contribution in \cite{F09}. Other modes not appearing in~\eqref{HlmExp} are of order $\mathcal{O}(1/c^8)$ and are thus neglected here. Finally, the modes for negative $m$ can be recovered using $\hat{H}_{\ell, -m}=(-1)^\ell \hat{H}_{\ell m}^*$ where the star notation refers to the complex conjugate operation.

The modes $(2,2)$, $(2,1)$, $(3,3)$ and $(3,1)$ were already known for the 3.5PN waveform and derived in \cite{FMBI12,FBI15,Jij3PN}. We recover these published results. Furthermore, the computation has been pushed 0.5PN higher for the other modes which completes the full waveform at the 3.5PN order. The overlaps agree with the existing literature~\cite{K07,BFIS08}, and notably with the modes derived for a test particle around a Schwarzschild black hole~\cite{TSasa94,TTS96,FI10,Pan11,Pan:2011gk}.


\subsection{Effective-One-Body factorized modes}\label{subsec:eob} 

In this section, we recall the conventions and definitions of the resummed form of the modes convenient for EOB application. Following Refs.~\cite{Damour:2007xr,Damour:2007yf,Pan11,DIN09}, we rewrite the PN-expanded waveform modes of Eqs.~\eqref{eq:hlmexpl}--\eqref{HlmExp} in a factorized, resummed form as\footnote{In this section, we pose $c=G=1$.}
\begin{equation}
\label{hlmFact}
h_{\ell m}^\text{F} = h_{\ell m}^{(N,\epsilon_p)}\hat{S}_\text{eff}^{(\epsilon_p)}T_{\ell m}e^{\di \delta_{\ell m}}f_{\ell m}\,,
\end{equation}
where $\epsilon_p$ is the parity of $\ell+m$: $\epsilon_p = 0$ if $\ell+m$ is even, and $\epsilon_p = 1$ if $\ell+m$ is odd. We see that the modes are factorized in five blocks. The first one, $h_{\ell m}^{(N,\epsilon_p)}$ is the leading PN order contribution. It is known analytically for each $\ell$ and $m$~\cite{Th80,K07}.

The second factor is the effective source term $\hat{S}_\text{eff}$ which, depending on $\epsilon_p$, is either the effective energy $E_\text{eff}$ or the orbital angular momentum $p_\phi$,
\begin{equation}
\hat{S}_\text{eff} = \left\{
        \begin{array}{ll}
            \frac{E_\text{eff}(v)}{\mu} , & \quad \ell + m \text{ even} \\
           v\, \frac{p_\phi(v)}{\mu\tmass}, & \quad \ell + m \text{ odd}
        \end{array}
    \right. ,
\end{equation}
where $\mu\equiv m_1m_2/\tmass$ is the reduced mass, $v\equiv\left(\tmass \omega\right)^{1/3}=\sqrt{x}$ and $E_\text{eff}$ is related to the total energy  $E$ via the EOB energy map $E = \tmass \sqrt{1+2\nu \left(E_\text{eff}/\mu - 1\right)}$.

Similarly to the phase redefinition \eqref{psi}, the factor $T_{\ell m}$  allows to absorb the ``leading logarithms'' induced by tail effects~\cite{Poisson:1993vp,Blanchet:1997jj,AF97}. It is given by
\begin{equation}
T_{\ell m} = \frac{\Gamma\left(\ell + 1 - 2 \di \hat{k}\right)}{\Gamma (\ell + 1)} e^{\pi \hat{k}} e^{2\di \hat{k} \ln (2m\omega r_0)},
\end{equation}
where $\Gamma(...)$ is the Euler gamma function, $\hat{k}\equiv m \omega E$ and the constant $r_0$, related to $\omega_0$, takes the value $2\tmass/\sqrt{e}$.

The remaining part of the factorized modes is expressed as an amplitude $f_{\ell m}$ and a phase $\delta_{\ell m}$, which are computed such that the expansion of $h_{\ell m}^\text{F}$ agrees with the PN-expanded modes in Eq.~\eqref{eq:hlmexpl}.
We can further resum the amplitude term by introducing $\rho_{\ell m} = (f_{\ell m})^{1/\ell}$ which improves the agreement with numerical-relativity waveforms \cite{DIN09,Pan11}.
For spinning binaries, the non-spinning and spin contributions are separated for the odd $m$ modes, such that
\begin{align}
f_{\ell m} = \left\{
        \begin{array}{ll}
           \rho_{\ell m}^\ell, & \quad m \text{ even} \\
           (\rho_{\ell m}^\text{NS})^\ell + f_{\ell m}^\text{S}, & \quad m \text{ odd}
        \end{array}
    \right. , 
\end{align}
where $\rho_{\ell m}^\text{NS}$ is the non-spinning part of $\rho_{\ell m}$, while $f_{\ell m}^\text{S}$ is the spinning part of $f_{\ell m}$.\\

The explicit results for $\rho_{\ell m}$, $f_{\ell m}$ and $\delta_{\ell m}$ are displayed in Appendix~\ref{app:eob}. The energy $E$ in $\delta_{\ell m}$ is replaced by the Hamiltonian in EOB waveform models.

As discussed in~\cite{HMK22}, we remind that an error in the literature \cite{DIN09,Pan:2011gk} has been found in the $\mathcal{O}(\nu v^5)$ term of $\delta_{21}$. The coefficient -493/42 in these papers, coming from the radiation reaction contribution, should read -25/2, as we see in Eq.~\eqref{eq:delta21}. We also provide new PN terms: although the (2,1) PN expanded mode was published in~\cite{Jij3PN}, its factorized form was not. We finally obtain new coefficients in the (3,2) and $\ell\geq 4$ modes. These expressions are available in a \emph{Mathematica} file in the Supplemental Materials~\cite{SuppMaterial} including spinning effects.

\section{Summary}

In this paper, we computed the full gravitational waveform for non-spinning binaries in quasi-circular motion to the 3.5PN order. To do so, we used the PN-MPM formalism to derive the radiative multipole moments. All the computational methods were already developed in previous works (except for the generalization to next-to-leading order of some integration formulas, displayed in Appendix \ref{app:memory}, in order to compute the memory terms).

The waveform modes have been written in the conventional PN expanded way as well as a factorized form convenient for EOB template building. We already derived in~\cite{HMK22} all spin effects in the full waveform to the 3.5PN order in the  quasi-circular and non-precessing case. Combining these results, we obtain the waveform modes to the 3.5PN order including non-spinning and spinning effects.

We recall that in~\cite{HMK22}, we found some mistakes in the literature, notably some spin coefficients $\mathcal{O}(v^6 \chi^2 \nu^2)$ in the (2,1) mode. We also corrected a non-spinning term in $\delta_{21}$. In the present work, we agree fully on the overlap with the literature, except for this discrepancy in $\delta_{21}$.

We provide in the Supplemental Material~\cite{SuppMaterial} the full expressions of the modes for non-spinning and spinning effects. The notations and details regarding the spinning part of the modes are given in~\cite{HMK22}.

\section*{Acknowledgments}
The author is extremely grateful to grateful to Mohammed Khalil and François Larrouturou  who double checked some results. The author is also grateful to Luc Blanchet and Guillaume Faye for interesting discussions.


\appendix


\section{Generalization of formulas to compute memory terms}\label{app:memory}

The memory integrals are of the form
\begin{equation}
\int_{-\infty}^{T_R} \dd \tau A_L(\tau) B_{K}(\tau), 
\end{equation} 
where $A_L$ and $B_{L}$ are time derivatives of mass-type or current-type STF multipole moments. To compute them, one needs to know the value of integrals of the type
\begin{equation}\label{eq:int}
\int_{-\infty}^{T_R} \dd \tau \frac{e^{i n \phi(\tau)}}{r^p(\tau)},
\end{equation}
where we recall that $\phi$ is the orbital phase, $r$ is the separation, $n$ is an integer and $p$ is an integer or half-integer. The memory integrals have been derived in the 3PN full non-spinning waveform paper \cite{BFIS08}. At this order, only the leading PN order is required. However, for the 3.5PN waveform, we need to generalize the formula to the next-to-leading order (NLO). To do so, we introduce the adiabatic parameter evaluated at retarded time $T_R$
\begin{equation}
\xi(T_R) \equiv \frac{1}{(T_c-T_R) \,\omega(T_R)},
\end{equation}
where $T_c$ is the coalescence time. This quantity is a small parameter of order $\mathcal{O}(c^{-5})$. At NLO, the adiabatic parameter reads
\begin{equation}
\xi(T_R) = \frac{256}{5}\nu x^{5/2}(T_R)\bigl[1+ \kappa \, x(T_R) + \mathcal{O}(x^2)\bigr],
\end{equation}
where $\kappa = -\tfrac{743}{252} -\tfrac{11}{3}\nu$. Next, we perform the following change of variable
\begin{equation}
y\equiv \frac{T_R-\tau}{T_c-T_R},
\end{equation}
and we can express the phase and the separation in terms of $y$ at the NLO
\begin{subequations}\label{eq:rphi}
\begin{align}
r(\tau) &= r(T_R)(1+y)^{1/4}\left[1 + \eta\, x(T_R)\left( (1+y)^{-1/4} -1 \right) \right],\\
\phi(\tau) &= \phi(T_R) - \frac{8}{5\xi(T_R)}\bigg\{ \left[ (1+y)^{5/8}-1 \right]\\
&  \qquad \qquad \qquad \qquad \qquad  + x(T_R) \Big[ \zeta \left[ (1+y)^{5/8}-1 \right]  +\lambda \left[ (1+y)^{3/8}-1 \right]  \Big] \bigg\},\nn
\end{align}
\end{subequations}
where $\eta = -\tfrac{1751}{1008} - \tfrac{7}{12}\nu$, $\zeta = -\tfrac{743}{672} - \tfrac{11}{8}\nu$ and $\lambda = \tfrac{3715}{2016} + \tfrac{55}{24}\nu$. After inserting these expressions in \eqref{eq:int}, performing the new change of variable $z=\tfrac{8}{5}[(1+y)^{5/8}-1]$ and truncating the integrand to the NLO, we find
\begin{equation}
\int_{-\infty}^{T_R} \dd \tau \frac{e^{\di n \phi(\tau)}}{r^p(\tau)}= \frac{e^{\di n \phi(T_R)}}{r^p(T_R)\,\omega(T_R)\,\xi(T_R)} \int_0^\infty \dd z \frac{e^{-\frac{\di n}{\xi}z}}{\left(1+\frac{5z}{8}\right)^{\frac{2p-3}{5}}}\left[1+x(T_R) f(z) + \mathcal{O}(x^2) \right],
\end{equation}
where $f$ is a function of $z$, straightforward to compute, but lengthy to display. Finally, by performing two integrations by part and using the fact that $\xi = \mathcal{O}(c^{-5})$, we find for $n\neq 0$
\begin{equation}
\int_{-\infty}^{T_R} \dd \tau \frac{e^{\di n \phi(\tau)}}{r^p(\tau)}= \frac{e^{\di n \phi(T_R)}}{\di \,n \,r^p(T_R)\,\omega(T_R)}\biggl[1-x(T_R)\underbrace{\left( \zeta + \frac{3}{5} \lambda \right)}_{=0} + \mathcal{O}(x^2) \biggr].
\end{equation}
Interestingly, the combination $\zeta + \frac{3}{5} \lambda$ vanishes, which implies that we do not need to take into account the 1PN correction in Eqs.~\eqref{eq:rphi}. Thus, the 1PN correction only comes from the one of $r^p(T_R)$ which leads to the final formula
\begin{equation}\label{eq:intm}
\int_{-\infty}^{T_R} \dd \tau \frac{e^{\di n \phi(\tau)}}{r^p(\tau)}= \frac{c^{2p-3}}{(G \tmass)^{p-1}}\frac{e^{\di n \phi(T_R)}}{\di \,n} x^{p-3/2}\left[1+p \left(1-\frac{\nu}{3} \right)x + \mathcal{O}(x^2) \right].
\end{equation}
This formula is valid for $n\neq 0$, but when computing the memory integrals, we also encounter integrals with $n=0$. The same kind of reasoning allows to compute
\begin{equation}\label{eq:intm0}
\int_{-\infty}^{T_R}  \frac{\dd \tau}{r^p(\tau)}= \frac{5}{64\nu}\frac{c^{2p-3}}{(G \tmass)^{p-1}} \frac{x^{p-4}}{p-4}\left\{1+\left[-\kappa + p\left( \frac{\eta}{p-3}+1-\frac{\nu}{3} \right)\right]x + \mathcal{O}(x^2) \right\}.
\end{equation}
Note that in this case, the 1PN correction of $r^p(\tau)$ plays a role in the result. In our computations, the case $p=3$ and $p=4$ do not appear. With \eqref{eq:intm} and \eqref{eq:intm0}, we have everything in hand to compute the memory integrals required for the full 3.5PN waveform.

\section{Explicit quantities of the factorized modes}\label{app:eob}
In this Appendix, we write the explicit expressions for the factorized modes (see Sec.~\ref{subsec:eob})
\begin{subequations}\label{eq:flm}
\begin{align}
f_{22}&=1 + v^2 \left(- \frac{43}{21} + \frac{55}{42} \nu\right) + v^4 \left(- \frac{536}{189} - \frac{6745}{1512} \nu + \frac{2047}{1512} \nu^2\right)\nn \\
& + v^6
\left(\frac{21428357}{727650} -  \frac{856}{105}\left( \gamma_E + \ln 4v\right)  + \left(- \frac{34625}{3696} + \frac{41}{96} \pi^2\right) \nu -  \frac{227875}{33264} \nu^2 + \frac{114635}{99792} \nu^3\right)\,,\\
f_{21}&=1 + v^2 \left(- \frac{59}{28} + \frac{23}{42} \nu\right) + v^4 \left(- \frac{5}{9} - \frac{269}{126} \nu + \frac{85}{252}\nu^2\right) \nn \\
&+ v^6 \left( \frac{88404893}{11642400} - \frac{214}{105}\left( \gamma_E +\ln 2v\right) + \left(\frac{86699}{66528} - \frac{41}{384} \pi^2\right) \nu - \frac{37241}{66528} \nu^2 + \frac{9365}{49896} \nu^3 \right) \,,\\
f_{33}&= 1 + v^2 \left(- \frac{7}{2} + 2 \nu\right) + v^4 \left(- \frac{443}{440} -  \frac{3401}{330} \nu + \frac{887}{330} \nu^2\right) \nn \\
& + v^6 \left(\frac{147471561}{2802800} -\frac{78}{7} \left(\gamma_E+\ln 6v\right) + \left(-\frac{17161}{2860} + \frac{41}{64} \pi^2\right) \nu -  \frac{27409}{1560}\nu^2 + \frac{8237}{2860} \nu^3 \right)\,,\\
f_{32}&= 1 + \frac{v^2}{1-3\nu}\left( -\frac{164}{45}+\frac{223}{18}\nu-\frac{32}{9}\nu^2\right) \nn \\
&+ \frac{v^4}{1-3\nu}\left(\frac{854}{495}-\frac{23443}{2376}\nu+\frac{31721}{1485}\nu^2-\frac{4943}{1485}\nu^3\right)\,,\\
f_{31}&=1 + v^2 \left(- \frac{13}{6} -  \frac{2}{3} \nu\right) + v^4 \left(\frac{1273}{792} - \frac{371}{198} \nu - \frac{247}{198}\nu^2\right)\nn \\
& + v^6 \left(\frac{400427563}{75675600} - \frac{26}{21}\left( \gamma_E +\ln 2v\right)+ \left(- \frac{788399}{77220} + \frac{41}{64} \pi^2\right) \nu + \frac{311225}{30888} \nu^2 -  \frac{17525}{15444} \nu^3 \right)\,,\\
f_{44}&= 1 + \frac{v^2}{1-3\nu}\left( -\frac{269}{55}+\frac{587}{33}\nu-\frac{175}{22}\nu^2\right) \nn \\
&+ \frac{v^4}{1-3\nu}\left(\frac{63002}{25025}-\frac{9163}{325}\nu+\frac{381541}{6435}\nu^2-\frac{226097}{17160}\nu^3 \right)\,,\\
f_{43}&=1 + \frac{v^2}{1-2\nu}\left( - \frac{111}{22} +\frac{547}{44} \nu -  \frac{40}{11} \nu^2\right) \nn \\
&+ \frac{v^4}{1-2\nu}\left(\frac{225543}{40040} -\frac{132529}{5720} \nu + \frac{100003}{3432} \nu^2 - \frac{3889}{858} \nu^3\right)\,,\\
f_{42}&=1  + \frac{v^2}{1-3\nu}\left( - \frac{191}{55} +\frac{353}{33} \nu - \frac{19}{22} \nu^2\right) \nn \\
&+ \frac{v^4}{1-3\nu}\left(\frac{76918}{25025} - \frac{53297}{3575} \nu + \frac{100552}{6435} \nu^2 + \frac{25783}{17160} \nu^3 \right)\,,\\
f_{41}&=1  + \frac{v^2}{1-2\nu}\left( - \frac{301}{66} +\frac{1385}{132} \nu -  \frac{24}{11} \nu^2\right) \nn \\
&+ \frac{v^4}{1-2\nu}\left(\frac{760181}{120120} - \frac{815329}{51480} \nu + \frac{163313}{10296} \nu^2 -\frac{1409}{858} \nu^3\right)\,,\\
f_{55}&=1 + \frac{v^2}{1-2\nu}\left( - \frac{487}{78} +\frac{649}{39} \nu - \frac{256}{39} \nu^2\right) \nn \\
&+ \frac{v^4}{1-2\nu}\left(\frac{50569}{6552} - \frac{39899}{819} \nu + \frac{4001}{63} \nu^2 - \frac{10567}{819} \nu^3\right)\,,\\
f_{54}&=1  + \frac{v^2}{1-5\nu+5\nu^2}\left( - \frac{2908}{455} +\frac{13717}{390} \nu -  \frac{1823}{39} \nu^2+\frac{476}{39}\nu^3\right)\,,\\
f_{53}&=1 + \frac{v^2}{1-2\nu}\left( - \frac{125}{26} +\frac{425}{39} \nu -  \frac{88}{39} \nu^2\right) \nn \\
&+ \frac{v^4}{1-2\nu}\left(\frac{69359}{10920} - \frac{34493}{1365} \nu + \frac{6299}{273} \nu^2 -  \frac{365}{273} \nu^3\right)\,,\\
f_{52}&=1 + \frac{v^2}{1-5\nu+5\nu^2}\left( - \frac{2638}{455} +\frac{12097}{390} \nu -  \frac{1499}{39} \nu^2+\frac{314}{39}\nu^3\right)\,,\\
f_{51}&=1 + \frac{v^2}{1-2\nu}\left( - \frac{319}{78} +\frac{313}{39} \nu -  \frac{4}{39} \nu^2\right) \nn \\
&+ \frac{v^4}{1-2\nu}\left(\frac{28859}{4680} - \frac{9673}{585} \nu + \frac{917}{117} \nu^2 + \frac{287}{117} \nu^3\right)\,,\\
f_{66}&=1 + \frac{v^2}{1-5\nu+5\nu^2}\left( - \frac{53}{7} +43\nu -  \frac{123}{2} \nu^2+\frac{39}{2}\nu^3\right)\,,\\
f_{65}&=1 + \frac{v^2}{1-4\nu+3\nu^2}\left( - \frac{185}{24} +\frac{419}{12} \nu - \frac{455}{12} \nu^2+\frac{55}{6}\nu^3\right)\,,\\
f_{64}&=1 + \frac{v^2}{1-5\nu+5\nu^2}\left( - \frac{43}{7} +33\nu -  \frac{83}{2} \nu^2+\frac{19}{2}\nu^3\right)\,,\\
f_{63}&=1 + \frac{v^2}{1-4\nu+3\nu^2}\left( - \frac{169}{24} +\frac{371}{12} \nu -  \frac{125}{4} \nu^2+\frac{13}{2}\nu^3\right)\,,\\
f_{62}&=1 + \frac{v^2}{1-5\nu+5\nu^2}\left( - \frac{37}{7} +27\nu -  \frac{59}{2} \nu^2+\frac{7}{2}\nu^3\right)\,,\\
f_{61}&=1 + \frac{v^2}{1-4\nu+3\nu^2}\left( - \frac{161}{24} +\frac{347}{12} \nu - \frac{335}{12} \nu^2+\frac{31}{6}\nu^3\right)\,,\\
f_{77}&=1  + \frac{v^2}{1-4\nu+3\nu^2}\left( - \frac{151}{17} +\frac{2123}{51} \nu - \frac{4963}{102} \nu^2+\frac{230}{17}\nu^3\right)\,,\\
f_{75}&=1  + \frac{v^2}{1-4\nu+3\nu^2}\left( - \frac{127}{17} +\frac{1691}{51} \nu - \frac{3523}{102} \nu^2+\frac{134}{17}\nu^3\right)\,,\\
f_{73}&=1  + \frac{v^2}{1-4\nu+3\nu^2}\left( - \frac{111}{17} +\frac{1403}{51} \nu - \frac{2563}{102} \nu^2+\frac{70}{17}\nu^3\right)\,,\\
f_{71}&=1  + \frac{v^2}{1-4\nu+3\nu^2}\left( - \frac{103}{17} +\frac{1259}{51} \nu - \frac{2083}{102} \nu^2+\frac{38}{17}\nu^3\right)\,,
\end{align}
\end{subequations}
and at the leading order, $f_{\ell m}=1$. Furthermore, the $\rho_{\ell m}$ read
\begin{subequations}\label{eq:rholm}
\begin{align}
\rho_{22}&=1 + v^2 \left(- \frac{43}{42} + \frac{55}{84} \nu\right) + v^4 \left(- \frac{20555}{10584} -  \frac{33025}{21168} \nu + \frac{19583}{42336} \nu^2\right)\nn \\
& + v^6
\left(\frac{1556919113}{122245200} -  \frac{428}{105}\left( \gamma_E +\ln 4v\right)
+ \left(- \frac{48993925}{9779616} + \frac{41}{192} \pi^2\right) \nu \nn \right. \\
& \left. \qquad\qquad -\frac{6292061}{3259872} \nu^2+ \frac{10620745}{39118464} \nu^3\right)\,,\\
\rho_{21}&=1 + v^2 \left(- \frac{59}{56} + \frac{23}{84} \nu\right) + v^4 \left(- \frac{47009}{56448} -  \frac{10993}{14112} \nu + \frac{617}{4704} \nu^2\right) \nn \\
&+ v^6 \left( \frac{7613184941}{2607897600} -  \frac{107}{105} \left(\gamma_E +\ln 2v\right) + \left( \frac{1024181}{17385984} -  \frac{41}{768} \pi^2\right) \nu  \nn \right. \\
& \left. \qquad\qquad +  \frac{622373}{8692992} \nu^2 + \frac{2266171}{39118464} \nu^3 \right) \,,\\
\rho_{33}&= 1 + v^2 \left(- \frac{7}{6} + \frac{2}{3} \nu\right) + v^4 \left(- \frac{6719}{3960} -  \frac{1861}{990} \nu + \frac{149}{330} \nu^2\right) \nn \\
& + v^6 \left(\frac{3203101567}{227026800} -  \frac{26}{7} \left(\gamma_E +\ln 6v\right) + \left(- \frac{129509}{25740} + \frac{41}{192} \pi^2\right) \nu \nn \right. \\
& \left. \qquad\qquad - \frac{274621}{154440} \nu^2 + \frac{12011}{46332} \nu^3 \right)\,,\\
\rho_{32}&= 1 + \frac{v^2}{1-3\nu}\left( -\frac{164}{135}+\frac{223}{54}\nu-\frac{32}{27}\nu^2\right) \nn \\
&+ \frac{v^4}{(1-3\nu)^2}\left(-\frac{180566}{200475}+\frac{1610009}{320760}\nu-\frac{945121}{320760}\nu^2-\frac{508474}{40095}\nu^3+\frac{77141}{40095}\nu^4 \right)\,,\\
\rho_{31}&=1 + v^2 \left(- \frac{13}{18} -  \frac{2}{9} \nu\right) + v^4 \left(\frac{101}{7128} -  \frac{1685}{1782} \nu -  \frac{829}{1782}\nu^2\right)\nn \\
& + v^6 \left(\frac{11706720301}{6129723600} -  \frac{26}{63} \left( \gamma_E +\ln 2v \right) + \left(- \frac{9688441}{2084940} + \frac{41}{192} \pi^2\right) \nu \nn \right. \\
& \left. \qquad\qquad + \frac{174535}{75816} \nu^2 -  \frac{727247}{1250964} \nu^3 \right)\,,\\
\rho_{44}&= 1 + \frac{v^2}{1-3\nu}\left( -\frac{269}{220}+\frac{587}{132}\nu-\frac{175}{88}\nu^2\right) \nn \\
&+ \frac{v^4}{(1-3\nu)^2}\left(-\frac{14210377}{8808800}+\frac{32485357}{4404400}\nu-\frac{1401149}{1415700}\nu^2-\frac{801565}{37752}\nu^3+\frac{3976393}{1006720}\nu^4 \right)\,,\\
\rho_{43}&=1  + \frac{v^2}{1-2\nu}\left( - \frac{111}{88} +\frac{547}{176} \nu -  \frac{10}{11} \nu^2\right) \nn \\
&+ \frac{v^4}{(1-2\nu)^2}\left(- \frac{6894273}{7047040} +\frac{22211989}{7047040} \nu + \frac{1032509}{1098240} \nu^2 -  \frac{181867}{25168} \nu^3 + \frac{19379}{18876} \nu^4\right)\,,\\
\rho_{42}&=1  + \frac{v^2}{1-3\nu}\left( - \frac{191}{220} +\frac{353}{132} \nu -  \frac{19}{88} \nu^2\right) \nn \\
&+ \frac{v^4}{(1-3\nu)^2}\left(- \frac{3190529}{8808800} + \frac{4108813}{4404400} \nu + \frac{21506941}{5662800} \nu^2 -  \frac{3628549}{377520} \nu^3 - \frac{1204847}{1006720} \nu^4\right)\,,\\
\rho_{41}&=1  + \frac{v^2}{1-2\nu}\left( - \frac{301}{264} +\frac{1385}{528} \nu -  \frac{6}{11} \nu^2\right) \nn \\
&+ \frac{v^4}{(1-2\nu)^2}\left(- \frac{7775491}{21141120} + \frac{117238909}{63423360} \nu - \frac{332099}{1098240} \nu^2 -\frac{917167}{226512} \nu^3 + \frac{7075}{18876} \nu^4\right)\,,\\
\rho_{55}&=1  + \frac{v^2}{1-2\nu}\left( - \frac{487}{390} +\frac{649}{195} \nu -  \frac{256}{195} \nu^2\right) \nn \\
&+ \frac{v^4}{(1-2\nu)^2}\left(- \frac{3353747}{2129400} + \frac{4038803}{1064700} \nu + \frac{925493}{266175} \nu^2 - \frac{932171}{88725} \nu^3 + \frac{456206}{266175} \nu^4\right)\,,\\
\rho_{54}&=1  + \frac{v^2}{1-5\nu+5\nu^2}\left( - \frac{2908}{2275} +\frac{13717}{1950} \nu -  \frac{1823}{195} \nu^2+\frac{476}{195}\nu^3\right)\,,\\
\rho_{53}&=1  + \frac{v^2}{1-2\nu}\left( - \frac{25}{26} +\frac{85}{39} \nu -  \frac{88}{195} \nu^2\right) \nn \\
&+ \frac{v^4}{(1-2\nu)^2}\left(- \frac{410833}{709800} + \frac{279697}{354900} \nu + \frac{928009}{266175} \nu^2 -  \frac{296117}{53235} \nu^3 + \frac{33934}{266175} \nu^4\right)\,,\\
\rho_{52}&=1  + \frac{v^2}{1-5\nu+5\nu^2}\left( - \frac{2638}{2275} +\frac{12097}{1950} \nu -  \frac{1499}{195} \nu^2+\frac{314}{195}\nu^3\right)\,,\\
\rho_{51}&=1  + \frac{v^2}{1-2\nu}\left( - \frac{319}{390} +\frac{313}{195} \nu -  \frac{4}{195} \nu^2\right) \nn \\
&+ \frac{v^4}{(1-2\nu)^2}\left(- \frac{31877}{304200} - \frac{79387}{152100} \nu + \frac{112613}{38025} \nu^2 -  \frac{31849}{12675} \nu^3 - \frac{37342}{38025} \nu^4\right)\,,\\
\rho_{66}&=1  + \frac{v^2}{1-5\nu+5\nu^2}\left( - \frac{53}{42} +\frac{43}{6} \nu -  \frac{41}{4} \nu^2+\frac{13}{4}\nu^3\right)\,,\\
\rho_{65}&=1  + \frac{v^2}{1-4\nu+3\nu^2}\left( - \frac{185}{144} +\frac{419}{72} \nu -  \frac{455}{72} \nu^2+\frac{55}{36}\nu^3\right)\,,\\
\rho_{64}&=1  + \frac{v^2}{1-5\nu+5\nu^2}\left( - \frac{43}{42} +\frac{11}{2} \nu -  \frac{83}{12} \nu^2+\frac{19}{12}\nu^3\right)\,,\\
\rho_{63}&=1  + \frac{v^2}{1-4\nu+3\nu^2}\left( - \frac{169}{144} +\frac{371}{72} \nu -  \frac{125}{24} \nu^2+\frac{13}{12}\nu^3\right)\,,\\
\rho_{62}&=1  + \frac{v^2}{1-5\nu+5\nu^2}\left( - \frac{37}{42} +\frac{9}{2} \nu -  \frac{59}{12} \nu^2+\frac{7}{12}\nu^3\right)\,,\\
\rho_{61}&=1  + \frac{v^2}{1-4\nu+3\nu^2}\left( - \frac{161}{144} +\frac{347}{72} \nu -  \frac{335}{72} \nu^2+\frac{31}{36}\nu^3\right)\,,\\
\rho_{77}&=1  + \frac{v^2}{1-4\nu+3\nu^2}\left( - \frac{151}{119} +\frac{2123}{357} \nu -  \frac{709}{102} \nu^2+\frac{230}{119}\nu^3\right)\,,\\
\rho_{75}&=1  + \frac{v^2}{1-4\nu+3\nu^2}\left( - \frac{127}{119} +\frac{1691}{357} \nu -  \frac{3523}{714} \nu^2+\frac{134}{119}\nu^3\right)\,,\\
\rho_{73}&=1  + \frac{v^2}{1-4\nu+3\nu^2}\left( - \frac{111}{119} +\frac{1403}{357} \nu - \frac{2563}{714} \nu^2+\frac{10}{17}\nu^3\right)\,,\\
\rho_{71}&=1  + \frac{v^2}{1-4\nu+3\nu^2}\left( - \frac{103}{119} +\frac{1259}{357} \nu - \frac{2083}{714} \nu^2+\frac{38}{119}\nu^3\right)\,.
\end{align}
\end{subequations}
As for the $f_{\ell m}$, at leading order $\rho_{\ell m}=1$. Finally, we have
\begin{subequations}
\begin{align}
\delta_{22}=& \frac{7}{3} \omega E -24\nu v^5+\frac{428\pi}{105}\left(\omega E\right)^2+v^7\nu \left(\frac{30995}{1134}+\frac{3662}{135}\nu\right)\,,\\
\delta_{21}=& \frac{2}{3} \omega E -\frac{25}{2}\nu v^5+\frac{107\pi}{105}\left(\omega E\right)^2\,,\label{eq:delta21}\\
\delta_{33}=& \frac{13}{10} \omega E -\frac{80897}{2430}\nu v^5+\frac{39\pi}{7}\left(\omega E\right)^2\,,\\
\delta_{32}=& \left( \frac{2}{3} +\frac{11}{5}\nu \right)\frac{\omega E}{1-3\nu} + \frac{v^5\nu}{(1-3\nu)^2}\left(-\frac{87893}{5400}+\frac{66871}{600}\nu-\frac{28691}{150}\nu^2 \right)\,,\\
\delta_{31}=& \frac{13}{30} \omega E -\frac{17}{10}\nu v^5+\frac{13\pi}{21}\left(\omega E\right)^2\,,\\
\delta_{44}=& \left( \frac{14}{15} +\frac{73}{40}\nu \right)\frac{\omega E}{1-3\nu} + \frac{v^5\nu}{(1-3\nu)^2}\left(-\frac{5794139}{140800}+\frac{35353757}{140800}\nu-\frac{2740989}{7040}\nu^2 \right)\,,\\
\delta_{43}=& \left( \frac{3}{5} +\frac{4961}{810}\nu \right)\frac{\omega E}{1-2\nu} \,,\\
\delta_{42}=& \left( \frac{7}{15} +\frac{14}{5}\nu \right)\frac{\omega E}{1-3\nu} + \frac{v^5\nu}{(1-3\nu)^2}\left(-\frac{125569}{4400}+\frac{764879}{4400}\nu-\frac{291003}{1100}\nu^2 \right)\,,\\
\delta_{41}=& \left( \frac{1}{5} +\frac{507}{10}\nu \right)\frac{\omega E}{1-2\nu} \,,\\
\delta_{55}=& \left( \frac{31}{42} +\frac{61252}{9375}\nu \right)\frac{\omega E}{1-2\nu} \,,\\
\delta_{54}=& \left( \frac{8}{15} +\frac{137377}{17920}\nu -\frac{11777}{384}\nu^2 \right)\frac{\omega E}{1-5\nu+5\nu^2} \,,\\
\delta_{53}=& \left( \frac{31}{70} +\frac{23924}{3645}\nu \right)\frac{\omega E}{1-2\nu} \,,\\
\delta_{52}=& \left( \frac{4}{15} +\frac{2351}{112}\nu -\frac{3943}{60}\nu^2 \right)\frac{\omega E}{1-5\nu+5\nu^2} \,,\\
\delta_{51}=& \left( \frac{31}{210} +\frac{1796}{15}\nu \right)\frac{\omega E}{1-2\nu} \,,\\
\delta_{66}=& \left( \frac{43}{70} +\frac{1756523}{217728}\nu -\frac{85439}{2592}\nu^2 \right)\frac{\omega E}{1-5\nu+5\nu^2} \,,\\
\delta_{64}=& \left( \frac{43}{105} +\frac{417559}{49152}\nu -\frac{393431}{12288}\nu^2 \right)\frac{\omega E}{1-5\nu+5\nu^2} \,,\\
\delta_{62}=& \left( \frac{43}{210} +\frac{387629}{13440}\nu -\frac{42859}{480}\nu^2 \right)\frac{\omega E}{1-5\nu+5\nu^2} \,,
\end{align}
\end{subequations}
while the values for the others $\delta_{\ell m}$ are 0 at the 3.5PN order.


\bibliography{ListeRef_H22}

\begin{thebibliography}{41}%
\makeatletter
\providecommand \@ifxundefined [1]{%
 \@ifx{#1\undefined}
}%
\providecommand \@ifnum [1]{%
 \ifnum #1\expandafter \@firstoftwo
 \else \expandafter \@secondoftwo
 \fi
}%
\providecommand \@ifx [1]{%
 \ifx #1\expandafter \@firstoftwo
 \else \expandafter \@secondoftwo
 \fi
}%
\providecommand \natexlab [1]{#1}%
\providecommand \enquote  [1]{``#1''}%
\providecommand \bibnamefont  [1]{#1}%
\providecommand \bibfnamefont [1]{#1}%
\providecommand \citenamefont [1]{#1}%
\providecommand \href@noop [0]{\@secondoftwo}%
\providecommand \href [0]{\begingroup \@sanitize@url \@href}%
\providecommand \@href[1]{\@@startlink{#1}\@@href}%
\providecommand \@@href[1]{\endgroup#1\@@endlink}%
\providecommand \@sanitize@url [0]{\catcode `\\12\catcode `\$12\catcode
  `\&12\catcode `\#12\catcode `\^12\catcode `\_12\catcode `\%12\relax}%
\providecommand \@@startlink[1]{}%
\providecommand \@@endlink[0]{}%
\providecommand \url  [0]{\begingroup\@sanitize@url \@url }%
\providecommand \@url [1]{\endgroup\@href {#1}{\urlprefix }}%
\providecommand \urlprefix  [0]{URL }%
\providecommand \Eprint [0]{\href }%
\providecommand \doibase [0]{https://doi.org/}%
\providecommand \selectlanguage [0]{\@gobble}%
\providecommand \bibinfo  [0]{\@secondoftwo}%
\providecommand \bibfield  [0]{\@secondoftwo}%
\providecommand \translation [1]{[#1]}%
\providecommand \BibitemOpen [0]{}%
\providecommand \bibitemStop [0]{}%
\providecommand \bibitemNoStop [0]{.\EOS\space}%
\providecommand \EOS [0]{\spacefactor3000\relax}%
\providecommand \BibitemShut  [1]{\csname bibitem#1\endcsname}%
\let\auto@bib@innerbib\@empty
\bibitem [{\citenamefont {Abbott}\ \emph {et~al.}(2019)\citenamefont {Abbott}
  \emph {et~al.}}]{LVCO1O2}%
  \BibitemOpen
  \bibfield  {author} {\bibinfo {author} {\bibfnamefont {B.~P.}\ \bibnamefont
  {Abbott}} \emph {et~al.} (\bibinfo {collaboration} {LIGO Scientific
  Collaboration and Virgo Collaboration}),\ }\bibfield  {title} {\bibinfo
  {title} {Gwtc-1: A gravitational-wave transient catalog of compact binary
  mergers observed by ligo and virgo during the first and second observing
  runs},\ }\href {https://doi.org/10.1103/PhysRevX.9.031040} {\bibfield
  {journal} {\bibinfo  {journal} {Phys. Rev. X}\ }\textbf {\bibinfo {volume}
  {9}},\ \bibinfo {pages} {031040} (\bibinfo {year} {2019})}\BibitemShut
  {NoStop}%
\bibitem [{\citenamefont {Abbott}\ \emph
  {et~al.}(2021{\natexlab{a}})\citenamefont {Abbott}, \citenamefont {Abbott},
  \citenamefont {Abraham}, \citenamefont {Acernese}, \citenamefont {Ackley},
  \citenamefont {Adams}, \citenamefont {Adams}, \citenamefont {Adhikari},
  \citenamefont {Adya}, \citenamefont {Affeldt},\ and\ \citenamefont
  {et~al.}}]{catalogO3a}%
  \BibitemOpen
  \bibfield  {author} {\bibinfo {author} {\bibfnamefont {R.}~\bibnamefont
  {Abbott}}, \bibinfo {author} {\bibfnamefont {T.}~\bibnamefont {Abbott}},
  \bibinfo {author} {\bibfnamefont {S.}~\bibnamefont {Abraham}}, \bibinfo
  {author} {\bibfnamefont {F.}~\bibnamefont {Acernese}}, \bibinfo {author}
  {\bibfnamefont {K.}~\bibnamefont {Ackley}}, \bibinfo {author} {\bibfnamefont
  {A.}~\bibnamefont {Adams}}, \bibinfo {author} {\bibfnamefont
  {C.}~\bibnamefont {Adams}}, \bibinfo {author} {\bibfnamefont
  {R.}~\bibnamefont {Adhikari}}, \bibinfo {author} {\bibfnamefont
  {V.}~\bibnamefont {Adya}}, \bibinfo {author} {\bibfnamefont {C.}~\bibnamefont
  {Affeldt}},\ and\ \bibinfo {author} {\bibnamefont {et~al.}},\ }\bibfield
  {title} {\bibinfo {title} {Gwtc-2: Compact binary coalescences observed by
  ligo and virgo during the first half of the third observing run},\ }\bibfield
   {journal} {\bibinfo  {journal} {Physical Review X}\ }\textbf {\bibinfo
  {volume} {11}},\ \href {https://doi.org/10.1103/physrevx.11.021053}
  {10.1103/physrevx.11.021053} (\bibinfo {year}
  {2021}{\natexlab{a}})\BibitemShut {NoStop}%
\bibitem [{\citenamefont {Abbott}\ \emph
  {et~al.}(2021{\natexlab{b}})\citenamefont {Abbott} \emph
  {et~al.}}]{catalogO3b}%
  \BibitemOpen
  \bibfield  {author} {\bibinfo {author} {\bibfnamefont {R.}~\bibnamefont
  {Abbott}} \emph {et~al.} (\bibinfo {collaboration} {LIGO Scientific, VIRGO,
  KAGRA}),\ }\bibfield  {title} {\bibinfo {title} {{GWTC-3: Compact Binary
  Coalescences Observed by LIGO and Virgo During the Second Part of the Third
  Observing Run}},\ }\href@noop {} {\  (\bibinfo {year}
  {2021}{\natexlab{b}})},\ \Eprint {https://arxiv.org/abs/2111.03606}
  {arXiv:2111.03606 [gr-qc]} \BibitemShut {NoStop}%
\bibitem [{\citenamefont {{Amaro-Seoane \emph{et al.}}}(2017)}]{LISA17}%
  \BibitemOpen
  \bibfield  {author} {\bibinfo {author} {\bibnamefont {{Amaro-Seoane \emph{et
  al.}}}},\ }\bibfield  {title} {\bibinfo {title} {{Laser Interferometer Space
  Antenna}},\ }\href@noop {} {\bibfield  {journal} {\bibinfo  {journal} {arXiv
  e-prints}\ ,\ \bibinfo {eid} {arXiv:1702.00786}} (\bibinfo {year} {2017})},\
  \Eprint {https://arxiv.org/abs/1702.00786} {arXiv:1702.00786 [astro-ph.IM]}
  \BibitemShut {NoStop}%
\bibitem [{\citenamefont {Punturo}\ \emph {et~al.}(2010)\citenamefont {Punturo}
  \emph {et~al.}}]{ET10}%
  \BibitemOpen
  \bibfield  {author} {\bibinfo {author} {\bibfnamefont {M.}~\bibnamefont
  {Punturo}} \emph {et~al.},\ }\bibfield  {title} {\bibinfo {title} {{The
  Einstein Telescope: A third-generation gravitational wave observatory}},\
  }\href {https://doi.org/10.1088/0264-9381/27/19/194002} {\bibfield  {journal}
  {\bibinfo  {journal} {Class. Quant. Grav.}\ }\textbf {\bibinfo {volume}
  {27}},\ \bibinfo {pages} {194002} (\bibinfo {year} {2010})}\BibitemShut
  {NoStop}%
\bibitem [{\citenamefont {Buonanno}\ and\ \citenamefont
  {Damour}(1999)}]{BDEOB99}%
  \BibitemOpen
  \bibfield  {author} {\bibinfo {author} {\bibfnamefont {A.}~\bibnamefont
  {Buonanno}}\ and\ \bibinfo {author} {\bibfnamefont {T.}~\bibnamefont
  {Damour}},\ }\bibfield  {title} {\bibinfo {title} {{Effective one-body
  approach to general relativistic two-body dynamics}},\ }\href
  {https://doi.org/10.1103/PhysRevD.59.084006} {\bibfield  {journal} {\bibinfo
  {journal} {Phys. Rev. D}\ }\textbf {\bibinfo {volume} {59}},\ \bibinfo
  {pages} {084006} (\bibinfo {year} {1999})},\ \Eprint
  {https://arxiv.org/abs/gr-qc/9811091} {arXiv:gr-qc/9811091} \BibitemShut
  {NoStop}%
\bibitem [{\citenamefont {Buonanno}\ and\ \citenamefont
  {Damour}(2000)}]{BuonD00}%
  \BibitemOpen
  \bibfield  {author} {\bibinfo {author} {\bibfnamefont {A.}~\bibnamefont
  {Buonanno}}\ and\ \bibinfo {author} {\bibfnamefont {T.}~\bibnamefont
  {Damour}},\ }\bibfield  {title} {\bibinfo {title} {Transition from inspiral
  to plunge in binary black hole coalescences},\ }\href@noop {} {\bibfield
  {journal} {\bibinfo  {journal} {Phys. Rev. D}\ }\textbf {\bibinfo {volume}
  {62}},\ \bibinfo {pages} {064015} (\bibinfo {year} {2000})},\ \Eprint
  {https://arxiv.org/abs/gr-qc/0001013} {gr-qc/0001013} \BibitemShut {NoStop}%
\bibitem [{\citenamefont {Blanchet}\ \emph {et~al.}(2002)\citenamefont
  {Blanchet}, \citenamefont {Faye}, \citenamefont {Iyer},\ and\ \citenamefont
  {Joguet}}]{BFIJ02}%
  \BibitemOpen
  \bibfield  {author} {\bibinfo {author} {\bibfnamefont {L.}~\bibnamefont
  {Blanchet}}, \bibinfo {author} {\bibfnamefont {G.}~\bibnamefont {Faye}},
  \bibinfo {author} {\bibfnamefont {B.~R.}\ \bibnamefont {Iyer}},\ and\
  \bibinfo {author} {\bibfnamefont {B.}~\bibnamefont {Joguet}},\ }\bibfield
  {title} {\bibinfo {title} {Gravitational-wave inspiral of compact binary
  systems to 7/2 post-newtonian order},\ }\href@noop {} {\bibfield  {journal}
  {\bibinfo  {journal} {Phys. Rev. D}\ }\textbf {\bibinfo {volume} {65}},\
  \bibinfo {pages} {061501(R)} (\bibinfo {year} {2002})},\ \bibinfo {note}
  {erratum \textit{Phys. Rev. D}, 71:129902(E), 2005},\ \Eprint
  {https://arxiv.org/abs/gr-qc/0105099} {gr-qc/0105099} \BibitemShut {NoStop}%
\bibitem [{\citenamefont {Blanchet}\ \emph
  {et~al.}(2004{\natexlab{a}})\citenamefont {Blanchet}, \citenamefont {Damour},
  \citenamefont {Esposito-Far{\`e}se},\ and\ \citenamefont {Iyer}}]{BDEI04}%
  \BibitemOpen
  \bibfield  {author} {\bibinfo {author} {\bibfnamefont {L.}~\bibnamefont
  {Blanchet}}, \bibinfo {author} {\bibfnamefont {T.}~\bibnamefont {Damour}},
  \bibinfo {author} {\bibfnamefont {G.}~\bibnamefont {Esposito-Far{\`e}se}},\
  and\ \bibinfo {author} {\bibfnamefont {B.~R.}\ \bibnamefont {Iyer}},\
  }\bibfield  {title} {\bibinfo {title} {Gravitational radiation from
  inspiralling compact binaries completed at the third post-newtonian order},\
  }\href@noop {} {\bibfield  {journal} {\bibinfo  {journal} {Phys. Rev. Lett.}\
  }\textbf {\bibinfo {volume} {93}},\ \bibinfo {pages} {091101} (\bibinfo
  {year} {2004}{\natexlab{a}})},\ \Eprint {https://arxiv.org/abs/gr-qc/0406012}
  {gr-qc/0406012} \BibitemShut {NoStop}%
\bibitem [{\citenamefont {Marchand}\ \emph {et~al.}(2020)\citenamefont
  {Marchand}, \citenamefont {Henry}, \citenamefont {Larrouturou}, \citenamefont
  {Marsat}, \citenamefont {Faye},\ and\ \citenamefont {Blanchet}}]{MHLMFB20}%
  \BibitemOpen
  \bibfield  {author} {\bibinfo {author} {\bibfnamefont {T.}~\bibnamefont
  {Marchand}}, \bibinfo {author} {\bibfnamefont {Q.}~\bibnamefont {Henry}},
  \bibinfo {author} {\bibfnamefont {F.}~\bibnamefont {Larrouturou}}, \bibinfo
  {author} {\bibfnamefont {S.}~\bibnamefont {Marsat}}, \bibinfo {author}
  {\bibfnamefont {G.}~\bibnamefont {Faye}},\ and\ \bibinfo {author}
  {\bibfnamefont {L.}~\bibnamefont {Blanchet}},\ }\bibfield  {title} {\bibinfo
  {title} {{The mass quadrupole moment of compact binary systems at the fourth
  post-Newtonian order}},\ }\href {https://doi.org/10.1088/1361-6382/ab9ce1}
  {\bibfield  {journal} {\bibinfo  {journal} {Class. Quant. Grav.}\ }\textbf
  {\bibinfo {volume} {37}},\ \bibinfo {pages} {215006} (\bibinfo {year}
  {2020})},\ \Eprint {https://arxiv.org/abs/2003.13672} {arXiv:2003.13672
  [gr-qc]} \BibitemShut {NoStop}%
\bibitem [{\citenamefont {Henry}\ \emph {et~al.}(2021)\citenamefont {Henry},
  \citenamefont {Faye},\ and\ \citenamefont {Blanchet}}]{Jij3PN}%
  \BibitemOpen
  \bibfield  {author} {\bibinfo {author} {\bibfnamefont {Q.}~\bibnamefont
  {Henry}}, \bibinfo {author} {\bibfnamefont {G.}~\bibnamefont {Faye}},\ and\
  \bibinfo {author} {\bibfnamefont {L.}~\bibnamefont {Blanchet}},\ }\bibfield
  {title} {\bibinfo {title} {{The current-type quadrupole moment and
  gravitational-wave mode (\ensuremath{\ell}, m) = (2, 1) of compact binary
  systems at the third post-Newtonian order}},\ }\href
  {https://doi.org/10.1088/1361-6382/ac1850} {\bibfield  {journal} {\bibinfo
  {journal} {Class. Quant. Grav.}\ }\textbf {\bibinfo {volume} {38}},\ \bibinfo
  {pages} {185004} (\bibinfo {year} {2021})},\ \Eprint
  {https://arxiv.org/abs/2105.10876} {arXiv:2105.10876 [gr-qc]} \BibitemShut
  {NoStop}%
\bibitem [{\citenamefont {Larrouturou}\ \emph
  {et~al.}(2022{\natexlab{a}})\citenamefont {Larrouturou}, \citenamefont
  {Henry}, \citenamefont {Blanchet},\ and\ \citenamefont {Faye}}]{DDRIR1}%
  \BibitemOpen
  \bibfield  {author} {\bibinfo {author} {\bibfnamefont {F.}~\bibnamefont
  {Larrouturou}}, \bibinfo {author} {\bibfnamefont {Q.}~\bibnamefont {Henry}},
  \bibinfo {author} {\bibfnamefont {L.}~\bibnamefont {Blanchet}},\ and\
  \bibinfo {author} {\bibfnamefont {G.}~\bibnamefont {Faye}},\ }\bibfield
  {title} {\bibinfo {title} {{The quadrupole moment of compact binaries to the
  fourth post-Newtonian order: I. Non-locality in time and infra-red
  divergencies}},\ }\href {https://doi.org/10.1088/1361-6382/ac5762} {\bibfield
   {journal} {\bibinfo  {journal} {Class. Quant. Grav.}\ }\textbf {\bibinfo
  {volume} {39}},\ \bibinfo {pages} {115007} (\bibinfo {year}
  {2022}{\natexlab{a}})},\ \Eprint {https://arxiv.org/abs/2110.02240}
  {arXiv:2110.02240 [gr-qc]} \BibitemShut {NoStop}%
\bibitem [{\citenamefont {Larrouturou}\ \emph
  {et~al.}(2022{\natexlab{b}})\citenamefont {Larrouturou}, \citenamefont
  {Blanchet}, \citenamefont {Henry},\ and\ \citenamefont {Faye}}]{DDRIR2}%
  \BibitemOpen
  \bibfield  {author} {\bibinfo {author} {\bibfnamefont {F.}~\bibnamefont
  {Larrouturou}}, \bibinfo {author} {\bibfnamefont {L.}~\bibnamefont
  {Blanchet}}, \bibinfo {author} {\bibfnamefont {Q.}~\bibnamefont {Henry}},\
  and\ \bibinfo {author} {\bibfnamefont {G.}~\bibnamefont {Faye}},\ }\bibfield
  {title} {\bibinfo {title} {{The quadrupole moment of compact binaries to the
  fourth post-Newtonian order: II. Dimensional regularization and
  renormalization}},\ }\href {https://doi.org/10.1088/1361-6382/ac5ba0}
  {\bibfield  {journal} {\bibinfo  {journal} {Class. Quant. Grav.}\ }\textbf
  {\bibinfo {volume} {39}},\ \bibinfo {pages} {115008} (\bibinfo {year}
  {2022}{\natexlab{b}})},\ \Eprint {https://arxiv.org/abs/2110.02243}
  {arXiv:2110.02243 [gr-qc]} \BibitemShut {NoStop}%
\bibitem [{\citenamefont {Blanchet}\ \emph {et~al.}(2022)\citenamefont
  {Blanchet}, \citenamefont {Faye},\ and\ \citenamefont {Larrouturou}}]{BFL22}%
  \BibitemOpen
  \bibfield  {author} {\bibinfo {author} {\bibfnamefont {L.}~\bibnamefont
  {Blanchet}}, \bibinfo {author} {\bibfnamefont {G.}~\bibnamefont {Faye}},\
  and\ \bibinfo {author} {\bibfnamefont {F.}~\bibnamefont {Larrouturou}},\
  }\bibfield  {title} {\bibinfo {title} {{The Quadrupole Moment of Compact
  Binaries to the Fourth post-Newtonian Order: From Source to Canonical
  Moment}},\ }\href@noop {} {\  (\bibinfo {year} {2022})},\ \Eprint
  {https://arxiv.org/abs/2204.11293} {arXiv:2204.11293 [gr-qc]} \BibitemShut
  {NoStop}%
\bibitem [{\citenamefont {Trestini}\ \emph {et~al.}(2022)\citenamefont
  {Trestini}, \citenamefont {Larrouturou},\ and\ \citenamefont
  {Blanchet}}]{TLB22}%
  \BibitemOpen
  \bibfield  {author} {\bibinfo {author} {\bibfnamefont {D.}~\bibnamefont
  {Trestini}}, \bibinfo {author} {\bibfnamefont {F.}~\bibnamefont
  {Larrouturou}},\ and\ \bibinfo {author} {\bibfnamefont {L.}~\bibnamefont
  {Blanchet}},\ }\bibfield  {title} {\bibinfo {title} {{The Quadrupole Moment
  of Compact Binaries to the Fourth post-Newtonian Order: Relating the Harmonic
  and Radiative Metrics}},\ }\href@noop {} {\  (\bibinfo {year} {2022})},\
  \Eprint {https://arxiv.org/abs/2209.02719} {arXiv:2209.02719 [gr-qc]}
  \BibitemShut {NoStop}%
\bibitem [{\citenamefont {Tagoshi}\ and\ \citenamefont
  {Sasaki}(1994)}]{TSasa94}%
  \BibitemOpen
  \bibfield  {author} {\bibinfo {author} {\bibfnamefont {H.}~\bibnamefont
  {Tagoshi}}\ and\ \bibinfo {author} {\bibfnamefont {M.}~\bibnamefont
  {Sasaki}},\ }\bibfield  {title} {\bibinfo {title} {Post-newtonian expansion
  of gravitational-waves from a particle in circular orbit around a
  schwarzschild black-hole},\ }\href@noop {} {\bibfield  {journal} {\bibinfo
  {journal} {Prog. Theor. Phys.}\ }\textbf {\bibinfo {volume} {92}},\ \bibinfo
  {pages} {745} (\bibinfo {year} {1994})},\ \Eprint
  {https://arxiv.org/abs/gr-qc/9405062} {gr-qc/9405062} \BibitemShut {NoStop}%
\bibitem [{\citenamefont {Tanaka}\ \emph {et~al.}(1996)\citenamefont {Tanaka},
  \citenamefont {Tagoshi},\ and\ \citenamefont {Sasaki}}]{TTS96}%
  \BibitemOpen
  \bibfield  {author} {\bibinfo {author} {\bibfnamefont {T.}~\bibnamefont
  {Tanaka}}, \bibinfo {author} {\bibfnamefont {H.}~\bibnamefont {Tagoshi}},\
  and\ \bibinfo {author} {\bibfnamefont {M.}~\bibnamefont {Sasaki}},\
  }\bibfield  {title} {\bibinfo {title} {Gravitational waves by a particle in
  circular orbit around a schwarzschild black hole: 5.5 {P}ost-{N}ewtonian
  formula},\ }\href@noop {} {\bibfield  {journal} {\bibinfo  {journal} {Prog.
  Theor. Phys.}\ }\textbf {\bibinfo {volume} {96}},\ \bibinfo {pages} {1087}
  (\bibinfo {year} {1996})},\ \Eprint {https://arxiv.org/abs/gr-qc/9701050}
  {gr-qc/9701050} \BibitemShut {NoStop}%
\bibitem [{\citenamefont {Fujita}\ and\ \citenamefont {Iyer}(2010)}]{FI10}%
  \BibitemOpen
  \bibfield  {author} {\bibinfo {author} {\bibfnamefont {R.}~\bibnamefont
  {Fujita}}\ and\ \bibinfo {author} {\bibfnamefont {B.}~\bibnamefont {Iyer}},\
  }\bibfield  {title} {\bibinfo {title} {Spherical harmonic modes of 5.5
  post-newtonian gravitational wave polarizations and associated factorized
  resummed waveforms for a particle in circular orbit around a schwarzschild
  black hole},\ }\href@noop {} {\bibfield  {journal} {\bibinfo  {journal}
  {Phys. Rev. D}\ }\textbf {\bibinfo {volume} {82}},\ \bibinfo {pages} {044051}
  (\bibinfo {year} {2010})},\ \Eprint {https://arxiv.org/abs/arXiv:1005.2266
  [gr-qc]} {arXiv:1005.2266 [gr-qc]} \BibitemShut {NoStop}%
\bibitem [{\citenamefont {Blanchet}\ \emph {et~al.}(2008)\citenamefont
  {Blanchet}, \citenamefont {Faye}, \citenamefont {Iyer},\ and\ \citenamefont
  {Sinha}}]{BFIS08}%
  \BibitemOpen
  \bibfield  {author} {\bibinfo {author} {\bibfnamefont {L.}~\bibnamefont
  {Blanchet}}, \bibinfo {author} {\bibfnamefont {G.}~\bibnamefont {Faye}},
  \bibinfo {author} {\bibfnamefont {B.~R.}\ \bibnamefont {Iyer}},\ and\
  \bibinfo {author} {\bibfnamefont {S.}~\bibnamefont {Sinha}},\ }\bibfield
  {title} {\bibinfo {title} {{The Third post-Newtonian gravitational wave
  polarisations and associated spherical harmonic modes for inspiralling
  compact binaries in quasi-circular orbits}},\ }\href
  {https://doi.org/10.1088/0264-9381/25/16/165003} {\bibfield  {journal}
  {\bibinfo  {journal} {Class. Quant. Grav.}\ }\textbf {\bibinfo {volume}
  {25}},\ \bibinfo {pages} {165003} (\bibinfo {year} {2008})},\ \bibinfo {note}
  {[Erratum: Class.Quant.Grav. 29, 239501 (2012)]},\ \Eprint
  {https://arxiv.org/abs/0802.1249} {arXiv:0802.1249 [gr-qc]} \BibitemShut
  {NoStop}%
\bibitem [{\citenamefont {Faye}\ \emph {et~al.}(2012)\citenamefont {Faye},
  \citenamefont {Marsat}, \citenamefont {Blanchet},\ and\ \citenamefont
  {Iyer}}]{FMBI12}%
  \BibitemOpen
  \bibfield  {author} {\bibinfo {author} {\bibfnamefont {G.}~\bibnamefont
  {Faye}}, \bibinfo {author} {\bibfnamefont {S.}~\bibnamefont {Marsat}},
  \bibinfo {author} {\bibfnamefont {L.}~\bibnamefont {Blanchet}},\ and\
  \bibinfo {author} {\bibfnamefont {B.~R.}\ \bibnamefont {Iyer}},\ }\bibfield
  {title} {\bibinfo {title} {The third and a half post-newtonian gravitational
  wave quadrupole mode for quasi-circular inspiralling compact binaries},\
  }\href@noop {} {\bibfield  {journal} {\bibinfo  {journal} {Class. Quant.
  Grav.}\ }\textbf {\bibinfo {volume} {29}},\ \bibinfo {pages} {175004}
  (\bibinfo {year} {2012})},\ \Eprint {https://arxiv.org/abs/arXiv:1204.1043}
  {arXiv:1204.1043} \BibitemShut {NoStop}%
\bibitem [{\citenamefont {Faye}\ \emph {et~al.}(2015)\citenamefont {Faye},
  \citenamefont {Blanchet},\ and\ \citenamefont {Iyer}}]{FBI15}%
  \BibitemOpen
  \bibfield  {author} {\bibinfo {author} {\bibfnamefont {G.}~\bibnamefont
  {Faye}}, \bibinfo {author} {\bibfnamefont {L.}~\bibnamefont {Blanchet}},\
  and\ \bibinfo {author} {\bibfnamefont {B.~R.}\ \bibnamefont {Iyer}},\
  }\bibfield  {title} {\bibinfo {title} {Non-linear multipole interactions and
  gravitational-wave octupole modes for inspiralling compact binaries to
  third-and-a-half post-newtonian order},\ }\href@noop {} {\bibfield  {journal}
  {\bibinfo  {journal} {Class. Quant. Grav.}\ }\textbf {\bibinfo {volume}
  {32}},\ \bibinfo {pages} {045016} (\bibinfo {year} {2015})},\ \Eprint
  {https://arxiv.org/abs/arXiv:1409.3546} {arXiv:1409.3546} \BibitemShut
  {NoStop}%
\bibitem [{\citenamefont {Favata}(2009)}]{F09}%
  \BibitemOpen
  \bibfield  {author} {\bibinfo {author} {\bibfnamefont {M.}~\bibnamefont
  {Favata}},\ }\bibfield  {title} {\bibinfo {title} {Post-newtonian corrections
  to the gravitational-wave memory for quasicircular, inspiralling compact
  binaries},\ }\href@noop {} {\bibfield  {journal} {\bibinfo  {journal} {Phys.
  Rev. D}\ }\textbf {\bibinfo {volume} {80}},\ \bibinfo {pages} {024002}
  (\bibinfo {year} {2009})},\ \Eprint {https://arxiv.org/abs/arXiv:0812.0069}
  {arXiv:0812.0069} \BibitemShut {NoStop}%
\bibitem [{\citenamefont {Damour}\ and\ \citenamefont
  {Nagar}(2008)}]{Damour:2007yf}%
  \BibitemOpen
  \bibfield  {author} {\bibinfo {author} {\bibfnamefont {T.}~\bibnamefont
  {Damour}}\ and\ \bibinfo {author} {\bibfnamefont {A.}~\bibnamefont {Nagar}},\
  }\bibfield  {title} {\bibinfo {title} {{Comparing Effective-One-Body
  gravitational waveforms to accurate numerical data}},\ }\href
  {https://doi.org/10.1103/PhysRevD.77.024043} {\bibfield  {journal} {\bibinfo
  {journal} {Phys. Rev.}\ }\textbf {\bibinfo {volume} {D77}},\ \bibinfo {pages}
  {024043} (\bibinfo {year} {2008})},\ \Eprint
  {https://arxiv.org/abs/0711.2628} {arXiv:0711.2628 [gr-qc]} \BibitemShut
  {NoStop}%
\bibitem [{\citenamefont {Damour}\ \emph {et~al.}(2009)\citenamefont {Damour},
  \citenamefont {Iyer},\ and\ \citenamefont {Nagar}}]{DIN09}%
  \BibitemOpen
  \bibfield  {author} {\bibinfo {author} {\bibfnamefont {T.}~\bibnamefont
  {Damour}}, \bibinfo {author} {\bibfnamefont {B.}~\bibnamefont {Iyer}},\ and\
  \bibinfo {author} {\bibfnamefont {A.}~\bibnamefont {Nagar}},\ }\bibfield
  {title} {\bibinfo {title} {Improved resummation of post-newtonian multipolar
  waveforms from circularized compact binaries},\ }\href@noop {} {\bibfield
  {journal} {\bibinfo  {journal} {Phys. Rev. D}\ }\textbf {\bibinfo {volume}
  {79}},\ \bibinfo {pages} {064004} (\bibinfo {year} {2009})},\ \Eprint
  {https://arxiv.org/abs/arXiv:0811.2069} {arXiv:0811.2069} \BibitemShut
  {NoStop}%
\bibitem [{\citenamefont {Pan}\ \emph {et~al.}(2011{\natexlab{a}})\citenamefont
  {Pan}, \citenamefont {Buonanno}, \citenamefont {Boyle}, \citenamefont
  {Buchman}, \citenamefont {Kidder}, \citenamefont {Pfeiffer},\ and\
  \citenamefont {Scheel}}]{Pan:2011gk}%
  \BibitemOpen
  \bibfield  {author} {\bibinfo {author} {\bibfnamefont {Y.}~\bibnamefont
  {Pan}}, \bibinfo {author} {\bibfnamefont {A.}~\bibnamefont {Buonanno}},
  \bibinfo {author} {\bibfnamefont {M.}~\bibnamefont {Boyle}}, \bibinfo
  {author} {\bibfnamefont {L.~T.}\ \bibnamefont {Buchman}}, \bibinfo {author}
  {\bibfnamefont {L.~E.}\ \bibnamefont {Kidder}}, \bibinfo {author}
  {\bibfnamefont {H.~P.}\ \bibnamefont {Pfeiffer}},\ and\ \bibinfo {author}
  {\bibfnamefont {M.~A.}\ \bibnamefont {Scheel}},\ }\bibfield  {title}
  {\bibinfo {title} {{Inspiral-merger-ringdown multipolar waveforms of
  nonspinning black-hole binaries using the effective-one-body formalism}},\
  }\href {https://doi.org/10.1103/PhysRevD.84.124052} {\bibfield  {journal}
  {\bibinfo  {journal} {Phys. Rev. D}\ }\textbf {\bibinfo {volume} {84}},\
  \bibinfo {pages} {124052} (\bibinfo {year} {2011}{\natexlab{a}})},\ \Eprint
  {https://arxiv.org/abs/1106.1021} {arXiv:1106.1021 [gr-qc]} \BibitemShut
  {NoStop}%
\bibitem [{\citenamefont {Henry}\ \emph {et~al.}(2022)\citenamefont {Henry},
  \citenamefont {Marsat},\ and\ \citenamefont {Khalil}}]{HMK22}%
  \BibitemOpen
  \bibfield  {author} {\bibinfo {author} {\bibfnamefont {Q.}~\bibnamefont
  {Henry}}, \bibinfo {author} {\bibfnamefont {S.}~\bibnamefont {Marsat}},\ and\
  \bibinfo {author} {\bibfnamefont {M.}~\bibnamefont {Khalil}},\ }\bibfield
  {title} {\bibinfo {title} {{Spin contributions to the gravitational-waveform
  modes for spin-aligned binaries at the 3.5PN order}},\ }\href@noop {} {\
  (\bibinfo {year} {2022})},\ \Eprint {https://arxiv.org/abs/2209.00374}
  {arXiv:2209.00374 [gr-qc]} \BibitemShut {NoStop}%
\bibitem [{Sup()}]{SuppMaterial}%
  \BibitemOpen
  \href@noop {} {}\bibinfo {note} {The ancillary files
  \texttt{modes\_PNexp\_full\_35PN.dat.m} and
  \texttt{factorized\_modes\_3.5PN.dat.m} contain the waveform modes with 3.5PN
  including non-spinning and spinning contributions, written in a PN expansion
  and in a factorized form.}\BibitemShut {Stop}%
\bibitem [{\citenamefont {Blanchet}(2014)}]{BlanchetLR}%
  \BibitemOpen
  \bibfield  {author} {\bibinfo {author} {\bibfnamefont {L.}~\bibnamefont
  {Blanchet}},\ }\bibfield  {title} {\bibinfo {title} {Gravitational radiation
  from post-newtonian sources and inspiralling compact binaries},\ }\href@noop
  {} {\bibfield  {journal} {\bibinfo  {journal} {Living Rev. Relativ.}\
  }\textbf {\bibinfo {volume} {17}},\ \bibinfo {pages} {2} (\bibinfo {year}
  {2014})},\ \Eprint {https://arxiv.org/abs/arXiv:1310.1528} {arXiv:1310.1528}
  \BibitemShut {NoStop}%
\bibitem [{\citenamefont {Thorne}(1980)}]{Th80}%
  \BibitemOpen
  \bibfield  {author} {\bibinfo {author} {\bibfnamefont {K.}~\bibnamefont
  {Thorne}},\ }\bibfield  {title} {\bibinfo {title} {Multipole expansions of
  gravitational radiation},\ }\href@noop {} {\bibfield  {journal} {\bibinfo
  {journal} {Rev. Mod. Phys.}\ }\textbf {\bibinfo {volume} {52}},\ \bibinfo
  {pages} {299} (\bibinfo {year} {1980})}\BibitemShut {NoStop}%
\bibitem [{\citenamefont {Kidder}(2008)}]{K07}%
  \BibitemOpen
  \bibfield  {author} {\bibinfo {author} {\bibfnamefont {L.}~\bibnamefont
  {Kidder}},\ }\bibfield  {title} {\bibinfo {title} {Using full information
  when computing modes of post-newtonian waveforms from inspiralling compact
  binaries in circular orbits},\ }\href@noop {} {\bibfield  {journal} {\bibinfo
   {journal} {Phys. Rev. D}\ }\textbf {\bibinfo {volume} {77}},\ \bibinfo
  {pages} {044016} (\bibinfo {year} {2008})},\ \Eprint
  {https://arxiv.org/abs/arXiv:0710.0614} {arXiv:0710.0614} \BibitemShut
  {NoStop}%
\bibitem [{\citenamefont {Pan}\ \emph {et~al.}(2011{\natexlab{b}})\citenamefont
  {Pan}, \citenamefont {Buonanno}, \citenamefont {Fujita}, \citenamefont
  {Racine},\ and\ \citenamefont {Tagoshi}}]{Pan11}%
  \BibitemOpen
  \bibfield  {author} {\bibinfo {author} {\bibfnamefont {Y.}~\bibnamefont
  {Pan}}, \bibinfo {author} {\bibfnamefont {A.}~\bibnamefont {Buonanno}},
  \bibinfo {author} {\bibfnamefont {R.}~\bibnamefont {Fujita}}, \bibinfo
  {author} {\bibfnamefont {E.}~\bibnamefont {Racine}},\ and\ \bibinfo {author}
  {\bibfnamefont {H.}~\bibnamefont {Tagoshi}},\ }\bibfield  {title} {\bibinfo
  {title} {{Post-Newtonian factorized multipolar waveforms for spinning,
  non-precessing black-hole binaries}},\ }\href
  {https://doi.org/10.1103/PhysRevD.83.064003} {\bibfield  {journal} {\bibinfo
  {journal} {Phys. Rev. D}\ }\textbf {\bibinfo {volume} {D}},\ \bibinfo {pages}
  {064003} (\bibinfo {year} {2011}{\natexlab{b}})},\ \Eprint
  {https://arxiv.org/abs/1006.0431} {arXiv:1006.0431 [gr-qc]} \BibitemShut
  {NoStop}%
\bibitem [{\citenamefont {Marchand}\ \emph {et~al.}(2018)\citenamefont
  {Marchand}, \citenamefont {Bernard}, \citenamefont {Blanchet},\ and\
  \citenamefont {Faye}}]{MBBF17}%
  \BibitemOpen
  \bibfield  {author} {\bibinfo {author} {\bibfnamefont {T.}~\bibnamefont
  {Marchand}}, \bibinfo {author} {\bibfnamefont {L.}~\bibnamefont {Bernard}},
  \bibinfo {author} {\bibfnamefont {L.}~\bibnamefont {Blanchet}},\ and\
  \bibinfo {author} {\bibfnamefont {G.}~\bibnamefont {Faye}},\ }\bibfield
  {title} {\bibinfo {title} {Ambiguity-free completion of the equations of
  motion of compact binary systems at the fourth post-newtonian order},\
  }\href@noop {} {\bibfield  {journal} {\bibinfo  {journal} {Phys. Rev. D}\
  }\textbf {\bibinfo {volume} {97}},\ \bibinfo {pages} {044023} (\bibinfo
  {year} {2018})},\ \Eprint {https://arxiv.org/abs/arXiv:1707.09289 [gr-qc]}
  {arXiv:1707.09289 [gr-qc]} \BibitemShut {NoStop}%
\bibitem [{\citenamefont {Blanchet}\ \emph
  {et~al.}(2004{\natexlab{b}})\citenamefont {Blanchet}, \citenamefont
  {Damour},\ and\ \citenamefont {Esposito-Far{\`e}se}}]{BDE04}%
  \BibitemOpen
  \bibfield  {author} {\bibinfo {author} {\bibfnamefont {L.}~\bibnamefont
  {Blanchet}}, \bibinfo {author} {\bibfnamefont {T.}~\bibnamefont {Damour}},\
  and\ \bibinfo {author} {\bibfnamefont {G.}~\bibnamefont
  {Esposito-Far{\`e}se}},\ }\bibfield  {title} {\bibinfo {title} {Dimensional
  regularization of the third post-newtonian dynamics of point particles in
  harmonic coordinates},\ }\href@noop {} {\bibfield  {journal} {\bibinfo
  {journal} {Phys. Rev. D}\ }\textbf {\bibinfo {volume} {69}},\ \bibinfo
  {pages} {124007} (\bibinfo {year} {2004}{\natexlab{b}})},\ \Eprint
  {https://arxiv.org/abs/gr-qc/0311052} {gr-qc/0311052} \BibitemShut {NoStop}%
\bibitem [{\citenamefont {Blanchet}\ \emph {et~al.}(2005)\citenamefont
  {Blanchet}, \citenamefont {Damour}, \citenamefont {Esposito-Far{\`e}se},\
  and\ \citenamefont {Iyer}}]{BDEI05dr}%
  \BibitemOpen
  \bibfield  {author} {\bibinfo {author} {\bibfnamefont {L.}~\bibnamefont
  {Blanchet}}, \bibinfo {author} {\bibfnamefont {T.}~\bibnamefont {Damour}},
  \bibinfo {author} {\bibfnamefont {G.}~\bibnamefont {Esposito-Far{\`e}se}},\
  and\ \bibinfo {author} {\bibfnamefont {B.~R.}\ \bibnamefont {Iyer}},\
  }\bibfield  {title} {\bibinfo {title} {Dimensional regularization of the
  third post-newtonian gravitational wave generation of two point masses},\
  }\href@noop {} {\bibfield  {journal} {\bibinfo  {journal} {Phys. Rev. D}\
  }\textbf {\bibinfo {volume} {71}},\ \bibinfo {pages} {124004} (\bibinfo
  {year} {2005})},\ \Eprint {https://arxiv.org/abs/gr-qc/0503044}
  {gr-qc/0503044} \BibitemShut {NoStop}%
\bibitem [{\citenamefont {Goldberger}\ and\ \citenamefont
  {Ross}(2010)}]{GRoss10}%
  \BibitemOpen
  \bibfield  {author} {\bibinfo {author} {\bibfnamefont {W.}~\bibnamefont
  {Goldberger}}\ and\ \bibinfo {author} {\bibfnamefont {A.}~\bibnamefont
  {Ross}},\ }\bibfield  {title} {\bibinfo {title} {Gravitational radiative
  corrections from effective field theory},\ }\href@noop {} {\bibfield
  {journal} {\bibinfo  {journal} {Phys. Rev. D}\ }\textbf {\bibinfo {volume}
  {81}},\ \bibinfo {pages} {124015} (\bibinfo {year} {2010})},\ \Eprint
  {https://arxiv.org/abs/arXiv:0912.4254 [gr-qc]} {arXiv:0912.4254 [gr-qc]}
  \BibitemShut {NoStop}%
\bibitem [{\citenamefont {Blanchet}\ \emph {et~al.}(1998)\citenamefont
  {Blanchet}, \citenamefont {Faye},\ and\ \citenamefont {Ponsot}}]{BFP98}%
  \BibitemOpen
  \bibfield  {author} {\bibinfo {author} {\bibfnamefont {L.}~\bibnamefont
  {Blanchet}}, \bibinfo {author} {\bibfnamefont {G.}~\bibnamefont {Faye}},\
  and\ \bibinfo {author} {\bibfnamefont {B.}~\bibnamefont {Ponsot}},\
  }\bibfield  {title} {\bibinfo {title} {Gravitational field and equations of
  motion of compact binaries to 5/2 post-newtonian order},\ }\href@noop {}
  {\bibfield  {journal} {\bibinfo  {journal} {Phys. Rev. D}\ }\textbf {\bibinfo
  {volume} {58}},\ \bibinfo {pages} {124002} (\bibinfo {year} {1998})},\
  \Eprint {https://arxiv.org/abs/gr-qc/9804079} {gr-qc/9804079} \BibitemShut
  {NoStop}%
\bibitem [{\citenamefont {Bernard}\ \emph {et~al.}(2018)\citenamefont
  {Bernard}, \citenamefont {Blanchet}, \citenamefont {Faye},\ and\
  \citenamefont {Marchand}}]{BBFM17}%
  \BibitemOpen
  \bibfield  {author} {\bibinfo {author} {\bibfnamefont {L.}~\bibnamefont
  {Bernard}}, \bibinfo {author} {\bibfnamefont {L.}~\bibnamefont {Blanchet}},
  \bibinfo {author} {\bibfnamefont {G.}~\bibnamefont {Faye}},\ and\ \bibinfo
  {author} {\bibfnamefont {T.}~\bibnamefont {Marchand}},\ }\bibfield  {title}
  {\bibinfo {title} {Center-of-mass equations of motion and conserved integrals
  of compact binary systems at the fourth post-newtonian order},\ }\href@noop
  {} {\bibfield  {journal} {\bibinfo  {journal} {Phys. Rev. D}\ }\textbf
  {\bibinfo {volume} {97}},\ \bibinfo {pages} {044037} (\bibinfo {year}
  {2018})},\ \Eprint {https://arxiv.org/abs/arXiv:1711.00283 [gr-qc]}
  {arXiv:1711.00283 [gr-qc]} \BibitemShut {NoStop}%
\bibitem [{\citenamefont {Damour}\ and\ \citenamefont
  {Nagar}(2007)}]{Damour:2007xr}%
  \BibitemOpen
  \bibfield  {author} {\bibinfo {author} {\bibfnamefont {T.}~\bibnamefont
  {Damour}}\ and\ \bibinfo {author} {\bibfnamefont {A.}~\bibnamefont {Nagar}},\
  }\bibfield  {title} {\bibinfo {title} {{Faithful effective-one-body waveforms
  of small-mass-ratio coalescing black-hole binaries}},\ }\href
  {https://doi.org/10.1103/PhysRevD.76.064028} {\bibfield  {journal} {\bibinfo
  {journal} {Phys. Rev.}\ }\textbf {\bibinfo {volume} {D76}},\ \bibinfo {pages}
  {064028} (\bibinfo {year} {2007})},\ \Eprint
  {https://arxiv.org/abs/0705.2519} {arXiv:0705.2519 [gr-qc]} \BibitemShut
  {NoStop}%
\bibitem [{\citenamefont {Poisson}(1993)}]{Poisson:1993vp}%
  \BibitemOpen
  \bibfield  {author} {\bibinfo {author} {\bibfnamefont {E.}~\bibnamefont
  {Poisson}},\ }\bibfield  {title} {\bibinfo {title} {{Gravitational radiation
  from a particle in circular orbit around a black hole. 1: Analytical results
  for the nonrotating case}},\ }\href
  {https://doi.org/10.1103/PhysRevD.47.1497} {\bibfield  {journal} {\bibinfo
  {journal} {Phys. Rev. D}\ }\textbf {\bibinfo {volume} {47}},\ \bibinfo
  {pages} {1497} (\bibinfo {year} {1993})}\BibitemShut {NoStop}%
\bibitem [{\citenamefont {Blanchet}(1998)}]{Blanchet:1997jj}%
  \BibitemOpen
  \bibfield  {author} {\bibinfo {author} {\bibfnamefont {L.}~\bibnamefont
  {Blanchet}},\ }\bibfield  {title} {\bibinfo {title} {{Gravitational wave
  tails of tails}},\ }\href {https://doi.org/10.1088/0264-9381/15/1/009}
  {\bibfield  {journal} {\bibinfo  {journal} {Class. Quant. Grav.}\ }\textbf
  {\bibinfo {volume} {15}},\ \bibinfo {pages} {113} (\bibinfo {year} {1998})},\
  \bibinfo {note} {[Erratum: Class.Quant.Grav. 22, 3381 (2005)]},\ \Eprint
  {https://arxiv.org/abs/gr-qc/9710038} {arXiv:gr-qc/9710038} \BibitemShut
  {NoStop}%
\bibitem [{\citenamefont {Asada}\ and\ \citenamefont {Futamase}(1997)}]{AF97}%
  \BibitemOpen
  \bibfield  {author} {\bibinfo {author} {\bibfnamefont {H.}~\bibnamefont
  {Asada}}\ and\ \bibinfo {author} {\bibfnamefont {T.}~\bibnamefont
  {Futamase}},\ }\bibfield  {title} {\bibinfo {title} {{Propagation of
  gravitational waves from slow motion sources in Coulomb type potential}},\
  }\href {https://doi.org/10.1103/PhysRevD.56.R6062} {\bibfield  {journal}
  {\bibinfo  {journal} {Phys. Rev.}\ }\textbf {\bibinfo {volume} {D56}},\
  \bibinfo {pages} {6062} (\bibinfo {year} {1997})},\ \Eprint
  {https://arxiv.org/abs/gr-qc/9711009} {arXiv:gr-qc/9711009 [gr-qc]}
  \BibitemShut {NoStop}%
\end{thebibliography}%

\end{document}